\documentclass[12pt]{article}
\usepackage[dvips]{color}
\usepackage{epsfig}
\usepackage{amsmath}
\usepackage{graphicx}

\begin{document}
\title{\bf Interaction between viscous varying modified cosmic Chaplygin gas and Tachyonic fluid}
\author{{J. Sadeghi\thanks{Email: pouriya@ipm.ir}\hspace{1mm} and H. Farahani\thanks{Email:
h.farahani@umz.ac.ir}\hspace{1mm}}\\
{\small {\em  Department of physics, Mazandaran University,
Babolsar, Iran}}\\
{\small {\em P .O .Box 47416-95447, Babolsar, Iran}} } \maketitle
\begin{abstract}
In this paper we study the interaction between the general form of
viscous varying modified cosmic Chaplygin gas and the Tachyon fluid
in the framework of Einstein gravity. We want to reconstruct the
Tachyon potential and total equation of state parameter graphically
by using numerical methods. In the presence of deceleration
parameter, the interaction between components becomes sign
changeable to explain different evolutionary eras in the universe.
We review the potential and total equation of state parameter in
Emergent, Intermediate and Logamediate scenarios of scale factor
numerically. Analysis of total equation of state parameter show
that, $\omega_{\small{tot}}<-1$ and $\omega_{\small{tot}}>-1$ imply
the phantom-like and quintessence-like behaviors respectively. we
have checked the effects of cosmic and viscosity elements on the
interaction process. Stability is checked in all the models by the
squared velocity of
sound.
{\bf Pacs Number:} 95.35.+d; 95.85.-e; 98.80.-k
\end{abstract}
\newpage
\tableofcontents
\newpage
\section{Introduction}
The most attractive subject in cosmology is the accelerating
expansion of the universe which is based on the recent astrophysical
data explaining the universe is spatially flat and an invisible
cosmic fluid called dark energy with a hugely negative pressure
which is responsible for this expansion. This type of matter
violates the strong energy condition, i.e.,$\rho+3p<0$. There are
various phenomenological models describing dark energy. Cosmological
constant is the simplest one which gave rise to the $\Lambda$-CDM
model, but it suffers from two critical problems; fine tuning and
coincidence. The first one relates the small value of cosmological
constant and the second problem caused because we live in an epoch
that the magnitude of dark energy and dark matter are comparable.
Other candidates of dark energy are for examples quintessence [1],
phantom [2], quintom [3], tachyon [4], holographic dark energy [5],
K-essence [6] and various models of Chaplygin gas. The simplest case
of this model based on Chaplygin equation of state [7] to describe
the lifting force on a wing of an air plane in aerodynamics. The
Chaplygin gas (CG) was not consistent with observational data
[8-11]. Therefore, an extension of CG model proposed [12, 13], which
is called generalized Chaplygin gas (GCG). However, observational
data ruled out such a proposal. Then, GCG extends to the modified
Chaplygin gas (MCG) [14]. There is still more extensions such as
generalized
cosmic Chaplygin gas (GCCG) [15], and modified cosmic Chaplygin gas [16].\\
On the other hand bulk viscosity plays an important role in the
evolution of the universe. The idea that Chaplygin gas may has
viscosity first proposed by the Ref. [17] and then developed by [18-21].\\
In the Ref. [22] a model of varying Chaplyagin gas which interact
with a Tachyonic matter considered in framework of general
relativity, and investigated the potential $V(\phi)$ and
$\omega_{\small{tot}}$ numerically. Also, in the Ref. [23] a model
of varying generalized Chaplyagin gas which interact with a
Tachyonic fluid considered and $V(\phi)$ of the Tachyonic fluid and $\omega_{\small{tot}}$ of the mixture investigated graphically.\\
In this paper we are going to extend Refs. [22] and [23] to complete
version of Chaplygin gas and work on the interaction between the
mixture of dark matter and dark energy to study their behaviors
during evolution of the universe. To achieve this purpose sign
changeable interaction is used to include all the eras from the
early time until the end. By numerical methods and graphs, the case
of varying and the presence of bulk viscosity and cosmic elements
are considered. Constants are fixed by imposing the constraint
$V\rightarrow0$, and
the $\omega_{\small{tot}}$ implies the quintessence or phantom behaviors.\\
We work in the general relativity framework and in FRW metric since
has a spatially flat homogeneous and isotropic universe that is
suitable for Chaplygin gas as a fluid in it and cause the
acceleration expansion of the universe.\\
This paper organized as the following. In next section we review
evolution of Chaplygin gas (CG) to viscous varying modified cosmic
Chaplygin gas (VIVAMCCG) and in section 3 we introduce field
equations include interaction. In section 4 we study potential
$V(\phi)$ of the Tachyonic fluid and $\omega_{\small{tot}}$ of the
mixture for general case of interaction and various models of scale
factor. In section 5 we investigate stability of our model and
finally in section 6 we give conclusion.

\section{Chaplygin gas}
In this section we consider perfect fluids as a simple cosmological
model and CG [24] is the simplest case to fill the expanding
universe. The reason to choose CG is based on the observational data
that the equation of state parameter for dark energy can be less
than -1. Phantom fields are the same but they have instabilities
[25]. The CG equation of state is given  by the following [26],
\begin{equation}\label{s1}
p=-\frac{B}{\rho},
\end{equation}
where $B$ is a positive constant. The CG is also important in
holography [4], string theory [27], and supersymmetry [28]. Equation
(1) is connected to string theory and can be achieved by the
D-branes Nambu-Goto action which is moving in a (d+2)-dimensional
space-time in the light-cone parametrization [29]. CG is the only
kind of fluid that accepts a supersymmetric generalization [30].
Another remarkable characteristics of CG is their Euler equations
with a large group of symmetry which cause their integrability. CG
was studied before [31] in the stabilization of branes [32] and
black hole bulks [33]. A negative constant $B$ described in wiggly
strings which causes an anti-Chaplygin state equation [34, 35]. CG
describes a transition from a decelerated cosmological expansion to
the present cosmic acceleration and perhaps submit a deformation of
$\Lambda$-CDM models. The inhomogeneous CG can combine dark energy
and dark matter and plays the unification role of them [36, 37]. For
studying more about CG see Refs. [36-39]. It is also possible to
study FRW cosmology of a universe filled with generalized Chaplygin
gas (GCG) with the following equation of state [40-42],
\begin{equation}\label{s2}
p=-\frac{B}{\rho^\alpha},
\end{equation}
with $0<\alpha\leq1$. The GCG is also interesting from holography
point of view [43]. As we can see the GCG is corresponding to almost
dust ($p=0$) at high density which is not agree completely with our
universe. Therefore, modified Chaplygin gas (MCG) with the following
equation of state introduced [44, 45].
\begin{equation}\label{s3}
p=\mu\rho-\frac{B}{\rho^\alpha},
\end{equation}
where $\mu$ is a positive constant. This model is more appropriate
choice to have constant negative pressure at low energy density and
high pressure at high energy density. The special case of
$\mu=\frac{1}{3}$ is the best fitted value to describe evolution of
the universe from radiation regime to the $\Lambda$-CDM regime.\\
The next extension performed by the Ref. [15] where the generalized
cosmic Chaplygin gas (GCCG) introduced by the following equation of
state,
\begin{equation}\label{s4}
p=-\frac{1}{\rho^\alpha}\left[\frac{B}{1+\omega}-1+(\rho^{1+\alpha}-\frac{B}{1+\omega}+1)^{-\omega}\right].
\end{equation}
This model can also extend to varying modified cosmic Chaplygin
(VAMCCG) gas with the following equation of state,
\begin{equation}\label{s5}
p=\mu\rho-\frac{1}{\rho^\alpha}\left[\frac{B(a)}{1+\omega}-1+(\rho^{1+\alpha}
-\frac{B(a)}{1+\omega}+1)^{-\omega}\right]
\end{equation}
where $\omega$ is corresponding cosmic parameter. Here, $B(a)$ is no
longer a constant and depend on scale factor via the following
relation,
\begin{equation}\label{s6}
B(a)=-\omega(t)B_{0}a^{-3(1+\omega(t))(1+\alpha)},
\end{equation}
with,
\begin{equation}\label{s7}
\omega(t)=\omega_{0}+\omega_{1}(\frac{t\dot{H}}{H}),
\end{equation}
It is also possible to include bulk viscosity,
\begin{equation}\label{s8}
p=\mu\rho-\frac{1}{\rho^\alpha}\left[\frac{B(a)}{1+\omega}-1+(\rho^{1+\alpha}
-\frac{B(a)}{1+\omega}+1)^{-\omega}\right]-3\varsigma H,
\end{equation}
where $\varsigma$ is the viscosity coefficient and $H=\dot{a}$/a is
the Hubble parameter. The equation of state (8) corresponds to
viscous varying modified cosmic Chaplygin gas (VIVAMCCG) which is
our interesting
case in this paper.\\
The total energy density and pressure relate to the mixture of
VIVAMCCG interacting with Tachyon fluid and we define,
\begin{equation}\label{s9}
\rho_{\small{tot}}=\rho_{\small{TF}}+\rho_{DE},
\end{equation}
\begin{equation}\label{s10}
p_{\small{tot}}=p_{\small{TF}}+p_{DE},
\end{equation}
where $\rho_{TF}$ and $p_{TF}$ denote energy density and pressure of
Tachyonic fluid respectively. Also, $\rho_{DE}$ and $p_{DE}$ are
dark energy density and pressure which given by the equation of
state (8).
\section{Field equations}
As we know the Friedmann-Robertson-Walker (FRW) universe in
four-dimensional space-time is described by the following metric,
\begin{equation}\label{s11}
ds^2=-dt^2+a(t)^2\left(\frac{dr^2}{1-kr^{2}}+r^{2}d\Omega^{2}\right),
\end{equation}
where $d\Omega^{2}=d\theta^{2}+\sin^{2}\theta d\phi^{2}$, and $a(t)$
represents the scale factor. The $\theta$ and $\phi$ parameters are
the usual azimuthal and polar angles of spherical coordinates, with
$0\leq\theta\leq\pi$ and $0\leq\phi<2\pi$. The coordinates ($t, r,
\theta, \phi$) are called co-moving coordinates. Also, constant $k$
denotes the curvature of the space. The curvature $k$ may be not
only positive, corresponding to real finite radius, but also zero or
negative, corresponding to infinite or imaginary radius. The
possibilities are called closed ($k=1$), flat ($k=0$), and open
($k=-1$). In that case the Einstein equation is given by,
\begin{equation}\label{s12}
R_{\mu\nu}-\frac{1}{2}g_{\mu\nu}R=T_{\mu\nu},
\end{equation}
where we assumed $c=1$, $8\pi G = 1$ and $\Lambda=0$, $k=0$.\\
By
using the above relations one can obtain the following field
equations,
\begin{equation}\label{s13}
H^{2}=\frac{\rho_{tot}}{3},
\end{equation}
and
\begin{equation}\label{s14}
{\frac{\ddot{a}}{a}=-\frac{1}{6}(\rho_{tot}+p_{tot})}.
\end{equation}
The energy-momentum conservation law obtained as the following,
\begin{equation}\label{s15}
\dot{\rho}_{tot}+3(\rho_{tot}+{p_{tot}})H=0,
\end{equation}
Now, for the interaction between dark energy and matter, splits the
equation (15) into the following equations,
\begin{equation}\label{s16}
\dot{\rho}_{DE}+3(\rho_{DE}+{p_{DE}})H=Q,
\end{equation}
where $\rho_{DE}$ is given by the relations (1) to (8) which is
$\rho_{VIVAMCCG}$ for our complete model, and,
\begin{equation}\label{s17}
\dot{\rho}_{TF}+3(\rho_{TF}+{p_{TF}})H=-Q,
\end{equation}
where $Q$ is interaction term. Then, the energy density and pressure
of a Tachyonic fluid will be,
\begin{equation}\label{s18}
\rho_{TF}=3H^2-\rho_{DE},
\end{equation}
and
\begin{equation}\label{s19}
p_{TF}=\frac{-Q-\dot{\rho}_{TF}}{3H}-\rho_{TF},
\end{equation}
where we used the equation (13) and (17). Then, the equation of
state parameter and potential of Tachyonic matter are given by [22,
23],
\begin{equation}\label{s20}
\omega_{TF}(t)=\frac{p_{TF}}{\rho_{TF}},
\end{equation}
\begin{equation}\label{s21}
V(\phi)=\sqrt{-\rho_{TF}p_{TF}},
\end{equation}
and $\omega_{tot}$ is the equation of state parameter of the mixture
which given by,
\begin{equation}\label{s22}
\omega_{tot}=\frac{p_{TF}+p_{DE}}{\rho_{TF}+\rho_{DE}}.
\end{equation}
In order to investigate potential (21) and the total equation of
state parameter (22) we use conservation equation (16) to obtain
$\rho_{DE}$. Then, we use the equation (13) to obtain $\rho_{TF}$.
Therefore, the equation (14) gives us $p_{TF}$, and hence $V(\phi)$
and $\omega_{tot}$ will be obtained.
\section{Sign changeable interaction and scale factors}
One of the ways to solve the cosmological coincidence problem is to
consider the interaction between the components on phenomenological
level. Generally, interaction could be considered as a function of
energy densities and their derivatives:
$Q(\rho_{i},\dot{\rho_{i}},...)$. Interactions considered as:
$Q=3Hb\rho_{m}$, where $b>0$ is a coupling constant. The general
form is $Q=3Hb\gamma\rho_{i}+\gamma\dot{\rho_{i}}$, where $i={TF,
DE, tot}$, and $\gamma$ is dimensionless constant [46].These kind of
interactions are either positive or negative and cannot change sign.
A sign-changeable interaction is of the following form,
\begin{equation}\label{s23}
Q=q(\gamma\rho+3bH\rho),
\end{equation}
where $q$ is the deceleration parameter which is given by,
\begin{equation}\label{s24}
q=-(1+\frac{\dot{H}}{H^{2}}).
\end{equation}
Now, by putting the sign-changeable interaction and the equation of
state for viscous varying modified cosmic Chaplygin gas (8) in the
equation (16), we get to a differential equation as the following,
\begin{equation}\label{s25}
\dot{\rho}+3\left[\rho+\mu\rho-\frac{1}{\rho^\alpha}\left[\frac{B(a)}{1+\omega}-1+(\rho^{1+\alpha}
-\frac{B(a)}{1+\omega}+1)^{-\omega}\right]-3\varsigma
H\right]H=q(\gamma\rho+3bH\rho),
\end{equation}
which can be solved numerically for the $\rho$ during the time. Then
$p_{TF}$ and $\rho_{TF}$ can be achieved by equations (18) and
(19). Here, the variables are fixed to $V_{TF}\rightarrow0$. \\
According to the accelerating expansion of the universe, we work on
three kinds of scenarios which are based on different eras in the
evolutionary process of the universe which all of them consist of a
kind of expanding exponential scale factor as the followings.
\subsection{Emergent scenario}
The universe in the emergent scenario has some interesting
properties as the following. The universe at large scale is
isotropic and homogeneous and there is no time-like singularity. The
universe is accelerating and may contains exotic matter such as CG.
The scale factor in this scenario is given by,
\begin{equation}\label{s26}
a(t)=a_{0}(B+e^{Kt})^m,
\end{equation}
where $a_{0}>0$, $K>0$, $B>0$, and $m>1$ [36].\\ First, we consider
the simplest case of modified cosmic Chaplygin gas interacting with
Tachyon fluid in emergent era and reconstruct the tachyon potential
 and total equation of state parameter numerically (Fig. 1).\\
By imposing the bulk viscous parameter $\varsigma\neq0$ and variable
parameter $B(a)$ in the equation of state, we have the VIVAMCCG,
which is the general form. Here we suppose various models and find
their behaviors in respect to their $\omega_{tot}$ and according to
the constraint $V_{TF}\rightarrow0$. Analysis of Tachyon potential
$V(\phi)$ and $\omega_{tot}$ shows that in all models for the
emergent scenario and during whole evolution of the universe from
the beginning to the end, the total equation of state parameter is
$\omega_{tot}<-1$ which means the phantom-like behavior (Figs.
2-7).This shows that the Emergent scenario is useful for describing
the universe because it is in agreement with the real data. In the
presence of cosmic element in the VIVAMCCG in comparison to the
VIVAMCG, the Tachyon potential has a faster decreasing in time. In
VIMCCG and VAMCCG in comparison with the case of MCCG, The presence
of viscosity and varying condition cause slower vanishing for the
potential.
\subsection{Intermediate scenario}
For the scale factor corresponding to intermediate scenario we have,
\begin{equation}\label{s27}
a(t)=e^{\lambda t^\beta},
\end{equation}
where $\lambda>0$ and $0<\beta<1$ [37, 38].\\
Numerical research on the Tachyon potential and total equation of
state parameter in this era with the intermediate scale factor
reveals the quintessence-like behavior which is based on the
$\omega_{tot}>-1$ (Figs. 8-13).\\
In this case, the Tachyon potential of the VIVAMCCG vanishes slower
than the VIVAMCG. It the VIMCG model, it takes the longest time to
$V_{TF}\rightarrow0$. Here, we found a different reaction and
$V_{TF}$ in the VIVAMCCG decreases slower by the cosmic effect.
\subsection{Logamediate scenario}
The Logamediate scenario of the universe is motivated by considering
a class of possible cosmological solutions with indefinite
expansion. In this model the scale factor showing the accelerating
expansion of the universe is given by,
\begin{equation}\label{s28}
a(t)=e^{x (\ln(t))^\beta},
\end{equation}
where $x>0$, and $\beta>1$ [37].\\
The same process for the Logamediate scale factor gives the Figs.
14-18 which are similar to the Intermediate case consistent with the
quintessence-like behavior. Here, VIVAMCCG, acts has a rather slower
decrease in the value of $V_{TF}$ than the VIVAMCG potential.
\section{Stability}
Now, we are interested to check the stability to find the best model
for explaining the accelerating expansion of the universe. Positive
and bounded squared sound velocity of Chaplygin gas is a remarkable
feature that is a non-trivial fact of fluids based on their negative
pressure
\begin{equation}\label{s28}
\upsilon_{s}^{2}=\frac{\partial P}{\partial \rho}.
\end{equation}
Here, we study the stability of these variety of models with respect
to the positive quantity for $\upsilon_{s}^{2}$. This is important
to mention that the parameters are fixed according to
$V_{TF}\rightarrow0$ in the interaction.
\subsection{Emergent scenario}
In this case, the models VAMCCG, VAMCG, VIMCG and VIVAMCG are stable
during the whole time and the model of MCCG has no stability here.
For the model of VIMCCG, there is no stability until $t=15$ and for
VIVAMCCG, as the general form, the stability is present from the
beginning but after the time $t=108$, no stability exits in the
model.
\subsection{Intermediate scenario}
In the general model of VIVAMCCG the stability stays only till early
moments $t=9$ and stability continues no more. That is like the
emergent scenario but with this difference here that the stability
vanishes so soon in comparison with the similar model in emergent
kind. The models VIMCG, VAMCG, VIMCCG and VIVAMCG are stable all the
time of evolution. In the VAMCCG the stability disappears very soon
at $t=3$ that is a distinct situation of this scenario.
\subsection{Logamediate scenario}
For VIVAMCCG in logamediate scenario, the stability holds in early
time before $t=18$ and exactly like the other scenarios during the
late time the model misses stability. In the other models as
VIVAMCG, VAMCCG and VAMCG the stability is always available during
the whole evolution of the universe in respect to the time. But in
the VIMCCG as the emergent similar case, the stability holds until
the time $t=432$.
\section{Conclusion}
In this paper we extend Ref. [22] and [23] to the case of viscous
varying modified cosmic Chaplygin gas and studied interaction with
Tachyonic fluid. We considered general form of interaction and three
different cases of the universe which are Emergent, Intermediate and
Logamediate scenarios. In the Emergent case, we found the
phantom-like behavior due to the $\omega_{tot}$ and
quintessence-like behaviors in the Intermediate and Logamediate era
is obvious. The comparison between the values of $\omega_{tot}$ in
the case of $MCCG$ and in the presence of viscosity shows a tiny
difference that is negligible. Also, The cosmic element causes no
comparable difference in the $\omega_{tot}$. By taking
$\varsigma=0.3$ the values of $V_{TF}$ and
$\omega_{tot}$ decreases a little.\\
The viscosity causes a little delay in the vanishing procedure of
the Tachyon potential and the cosmic element appears by the same
feature.\\
Stability of the general model of VIVAMCCG of all the three
scenarios are contemporary during the time and vanishes in the late
time of expanding evolution of the universe. It shows that the
viscous varying modified cosmic Chaplygin gas model is valid for the
early universe.\\
Since the Phantom-like behavior is more consistent with the
observational data, Then the Emergent scenario is better than the
Logamediate and Intermediate ones and the stable models in this
scenario are the best ones among all the cases we have considered
here. These suitable models are VAMCCG, VAMCG, VIMCG and VIVAMCG in
Emergent scenario.

\begin{figure}[th]
\begin{center}
\includegraphics[scale=.4]{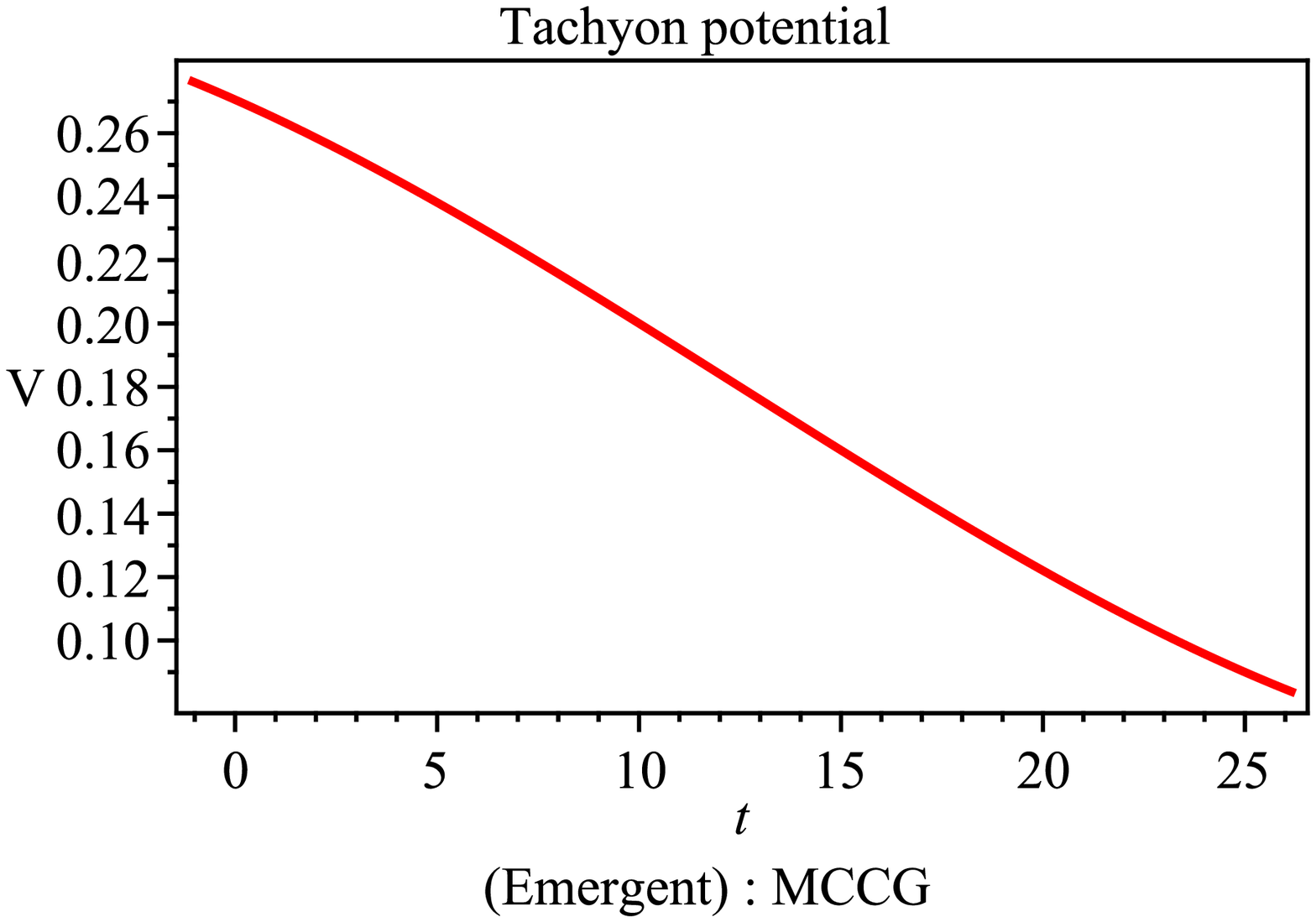}\includegraphics[scale=.4]{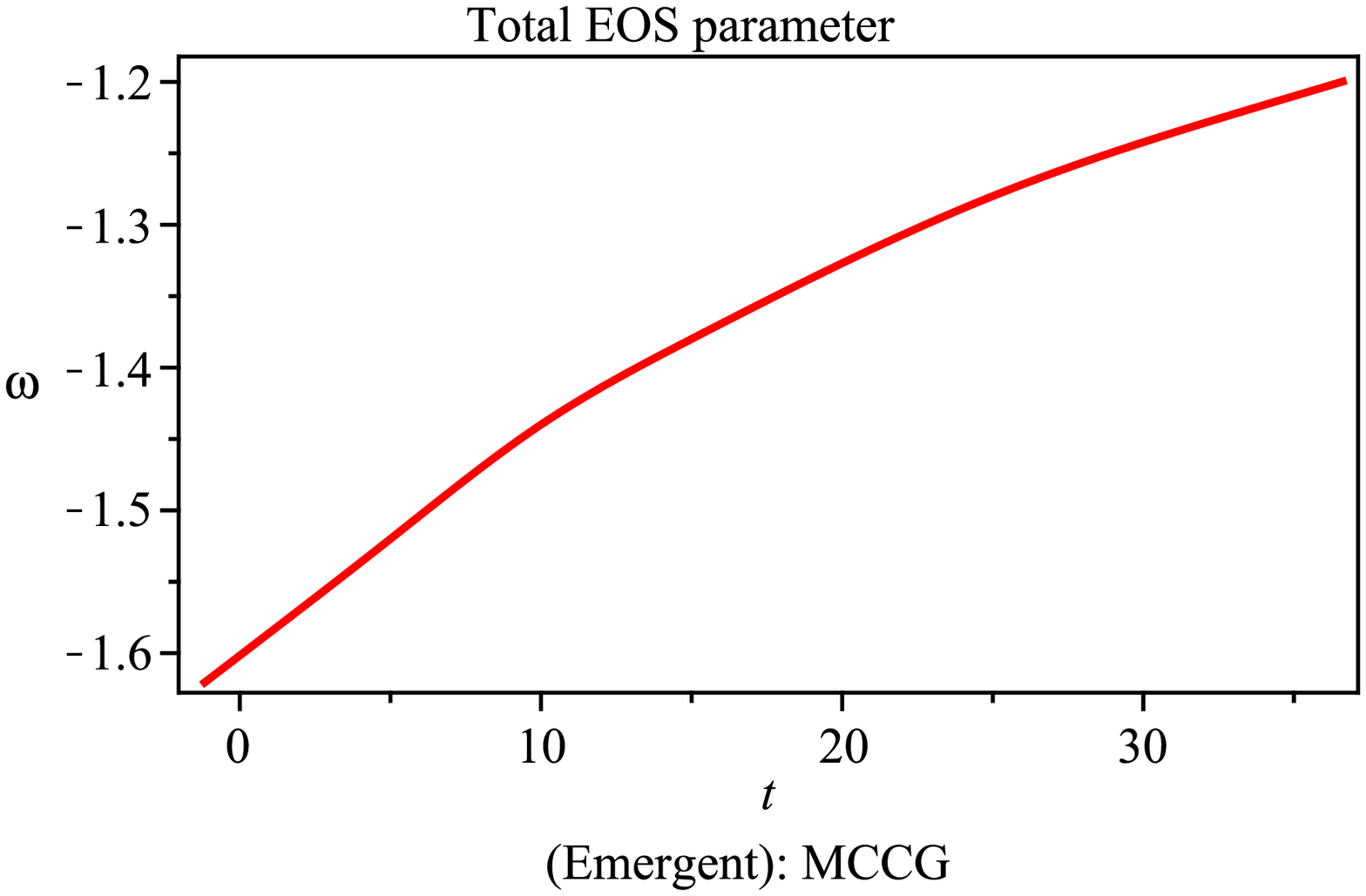}
\caption{Plot of $V$ and $\omega_{tot}$ in terms of time with
$m=1.1$, $A=-0.03$, $b=1$, $\gamma=0.7$, $B=1$, $a_{0}=0.6$,
$\mu=0.3$, $\omega=-0.5$, $\alpha=0.5$ and $K=0.03$.}
\end{center}
\end{figure}

\begin{figure}[th]
\begin{center}
\includegraphics[scale=.4]{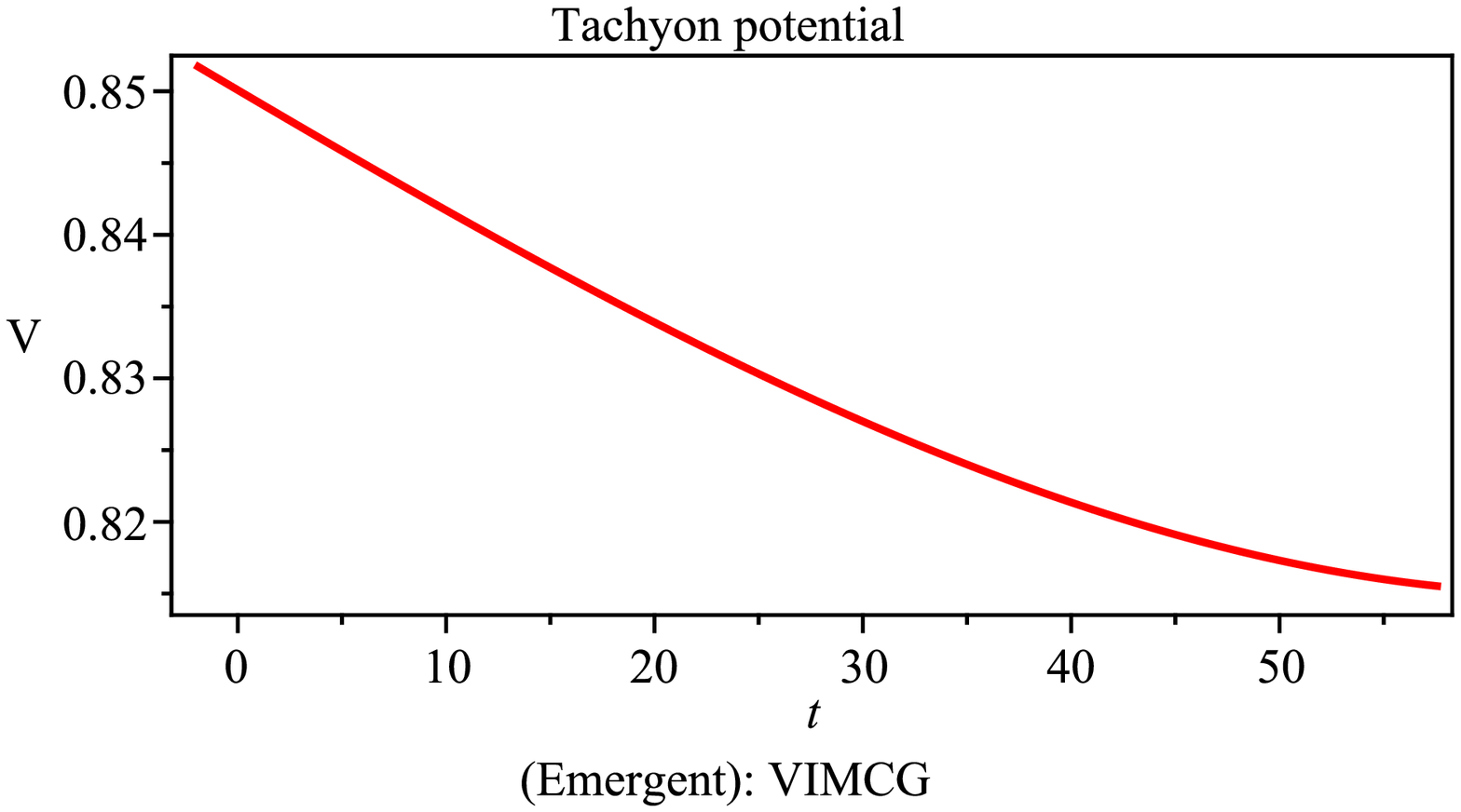}\includegraphics[scale=.4]{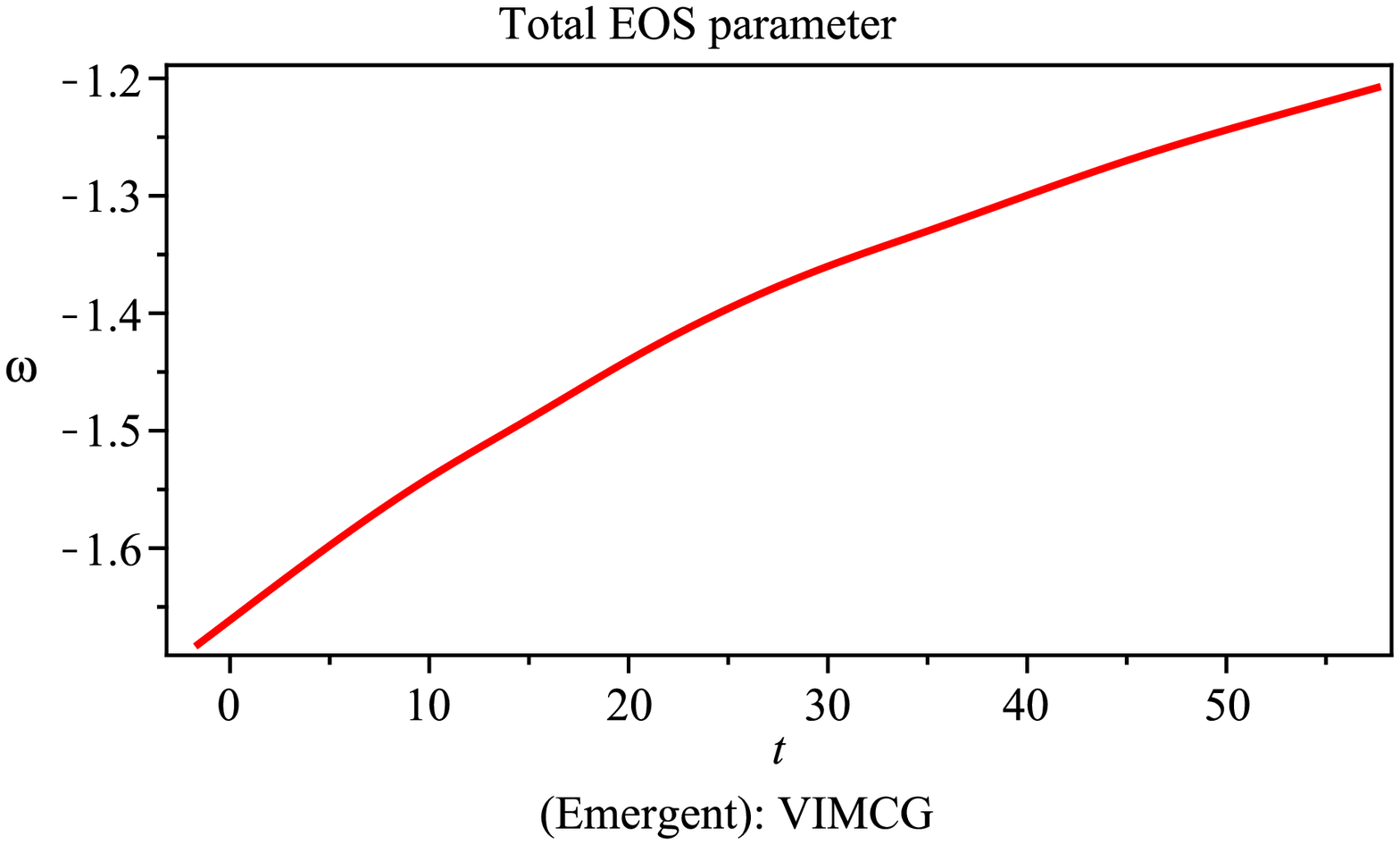}
\caption{Plot of $V$ and $\omega_{tot}$ in terms of time with $m=1$,
$A=1$, $b=1$, $\gamma=1$, $B=1$, $a_{0}=1$, $\mu=0.3$,
$\varsigma=1$, $\alpha=0.5$ and $K=0.02$.}
\end{center}
\end{figure}

\begin{figure}[th]
\begin{center}
\includegraphics[scale=.4]{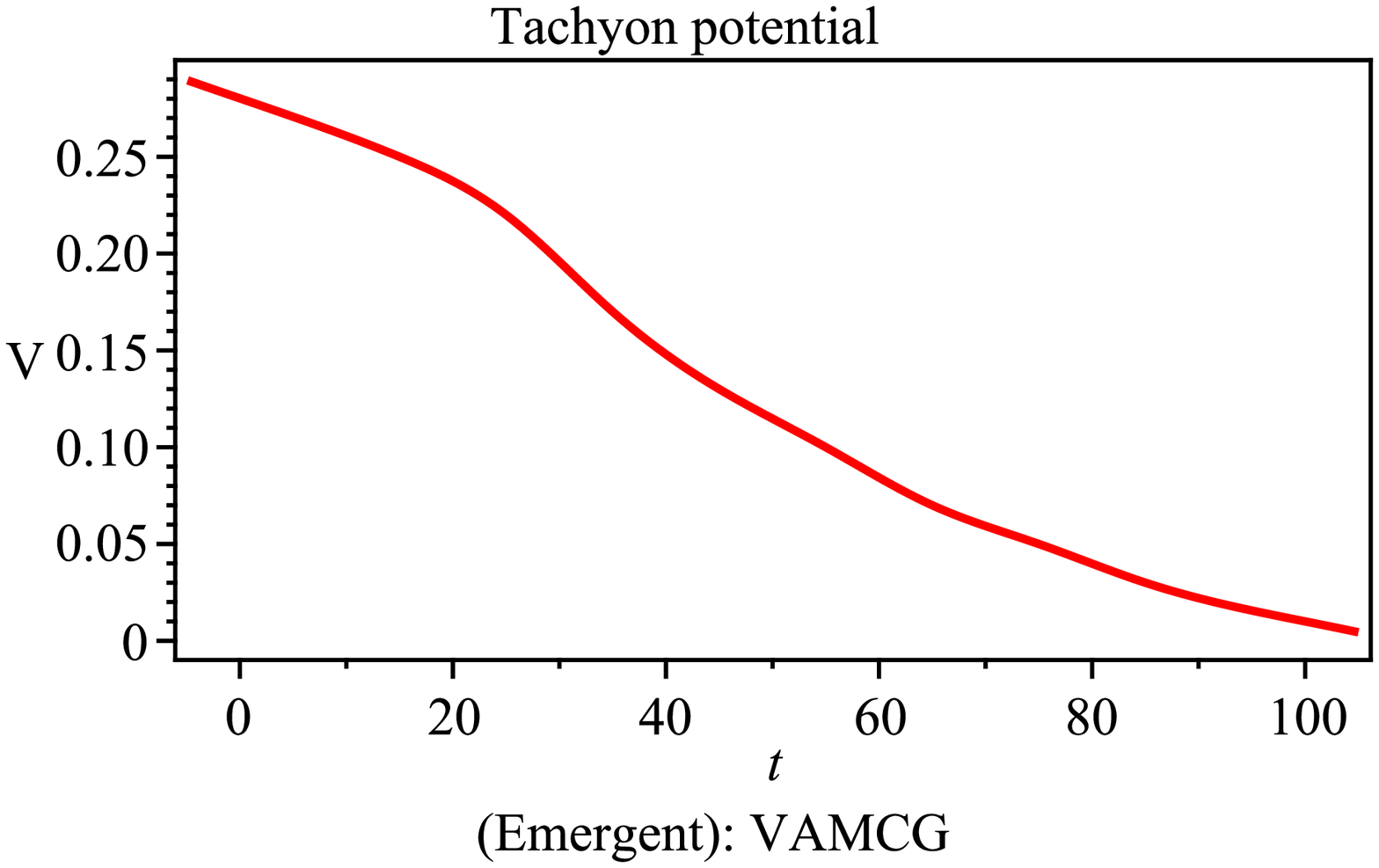}\includegraphics[scale=.4]{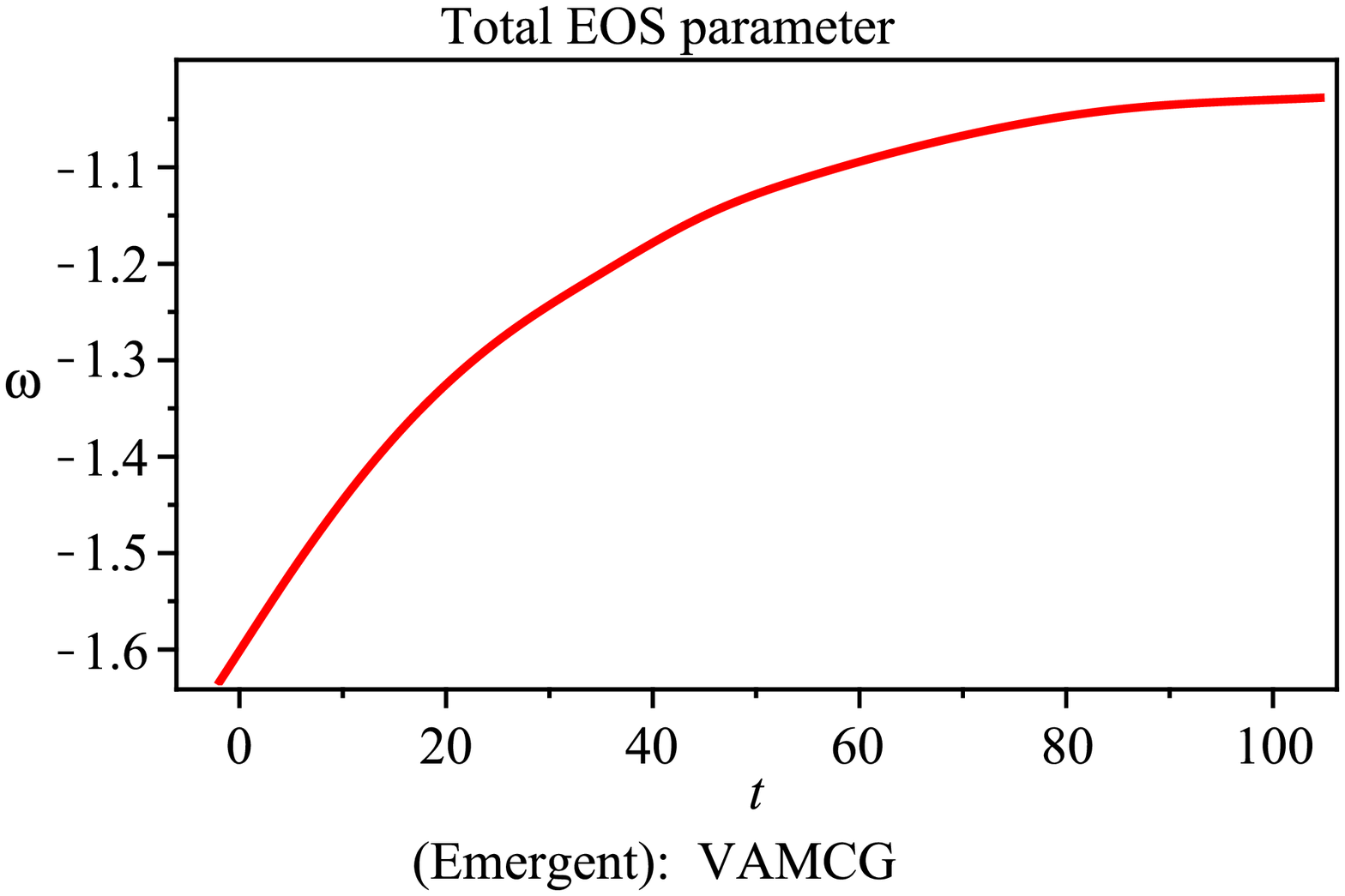}
\caption{Plot of $V$ and $\omega_{tot}$ in terms of time with
$m=1.1$, $b=0.1$, $\gamma=0.7$, $B=1$, $a_{0}=0.6$, $i=-0.6$,
$j=-0.08$, $\mu=0.3$, $\alpha=0.5$ and $K=0.03$.}
\end{center}
\end{figure}

\begin{figure}[th]
\begin{center}
\includegraphics[scale=.4]{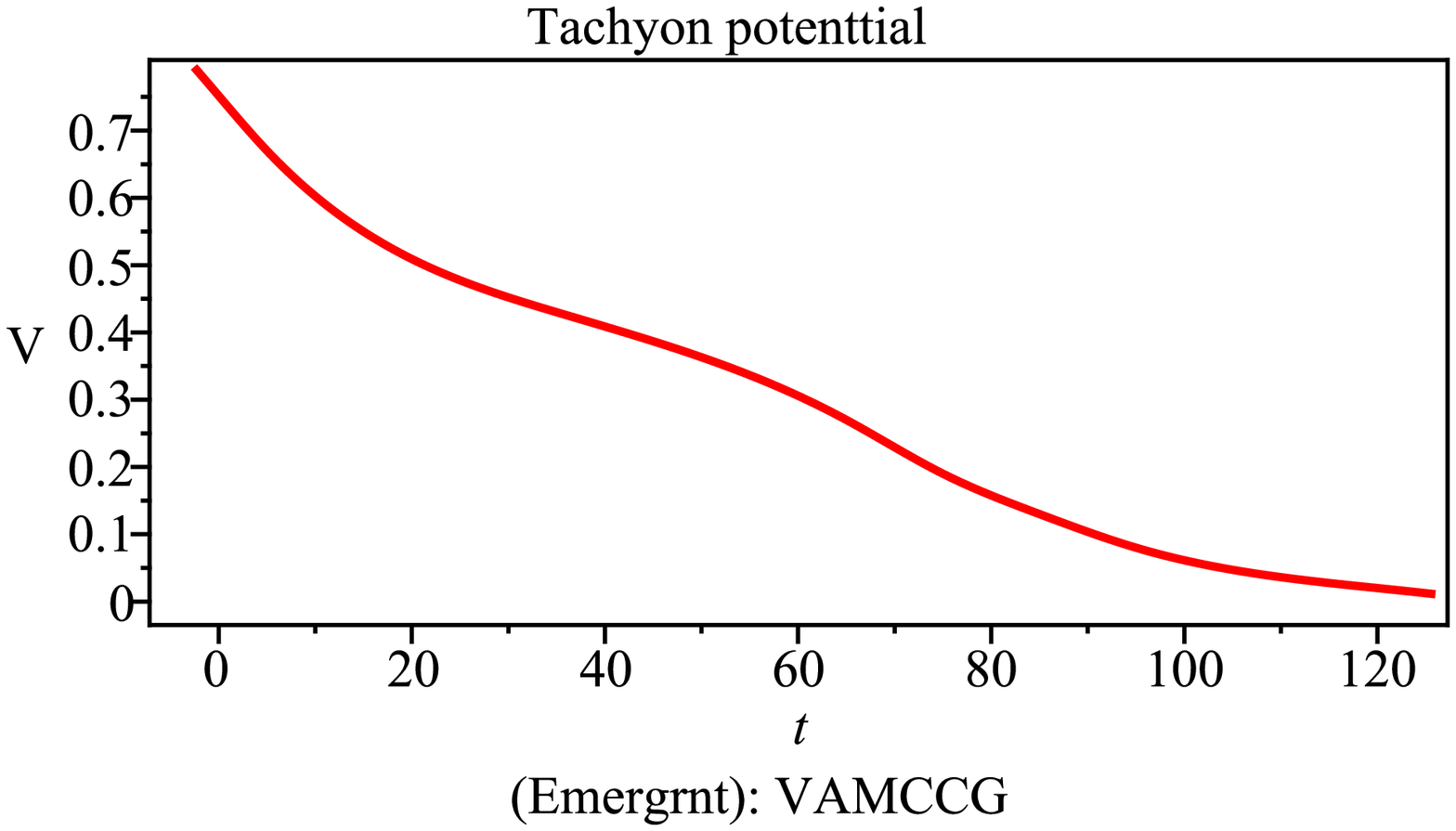}\includegraphics[scale=.4]{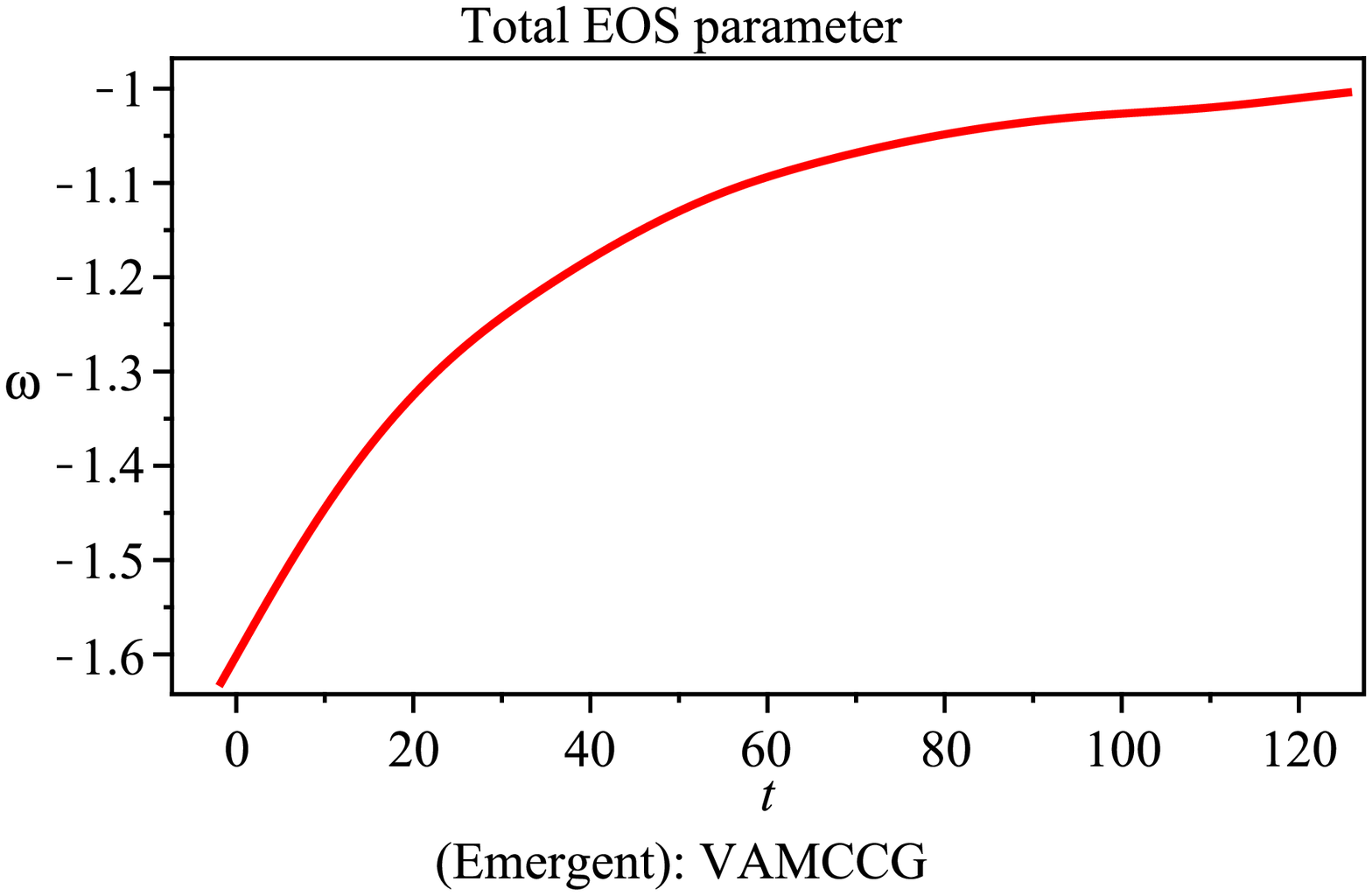}
\caption{Plot of $V$ and $\omega_{tot}$ in terms of time with
$m=1.1$, $b=1$, $K=0.03$, $\gamma=0.7$, $B=1$, $a_{0}=0.6$,
$i=-0.6$, $j=-0.8$, $\mu=0.3$, $\omega=-0.5$ $\alpha=0.5$ and
$K=0.03$.}
\end{center}
\end{figure}

\begin{figure}[th]
\begin{center}
\includegraphics[scale=.4]{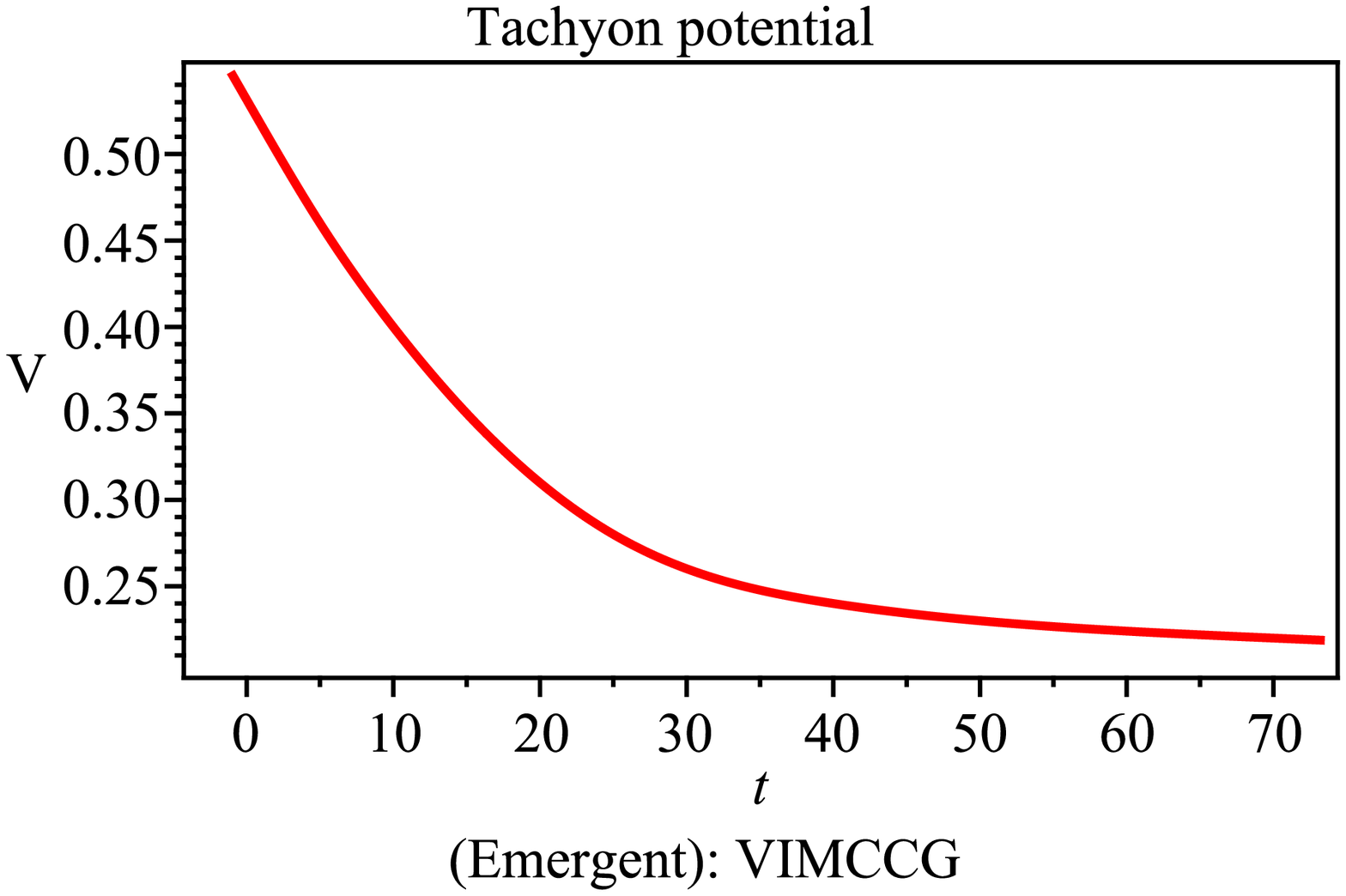}\includegraphics[scale=.4]{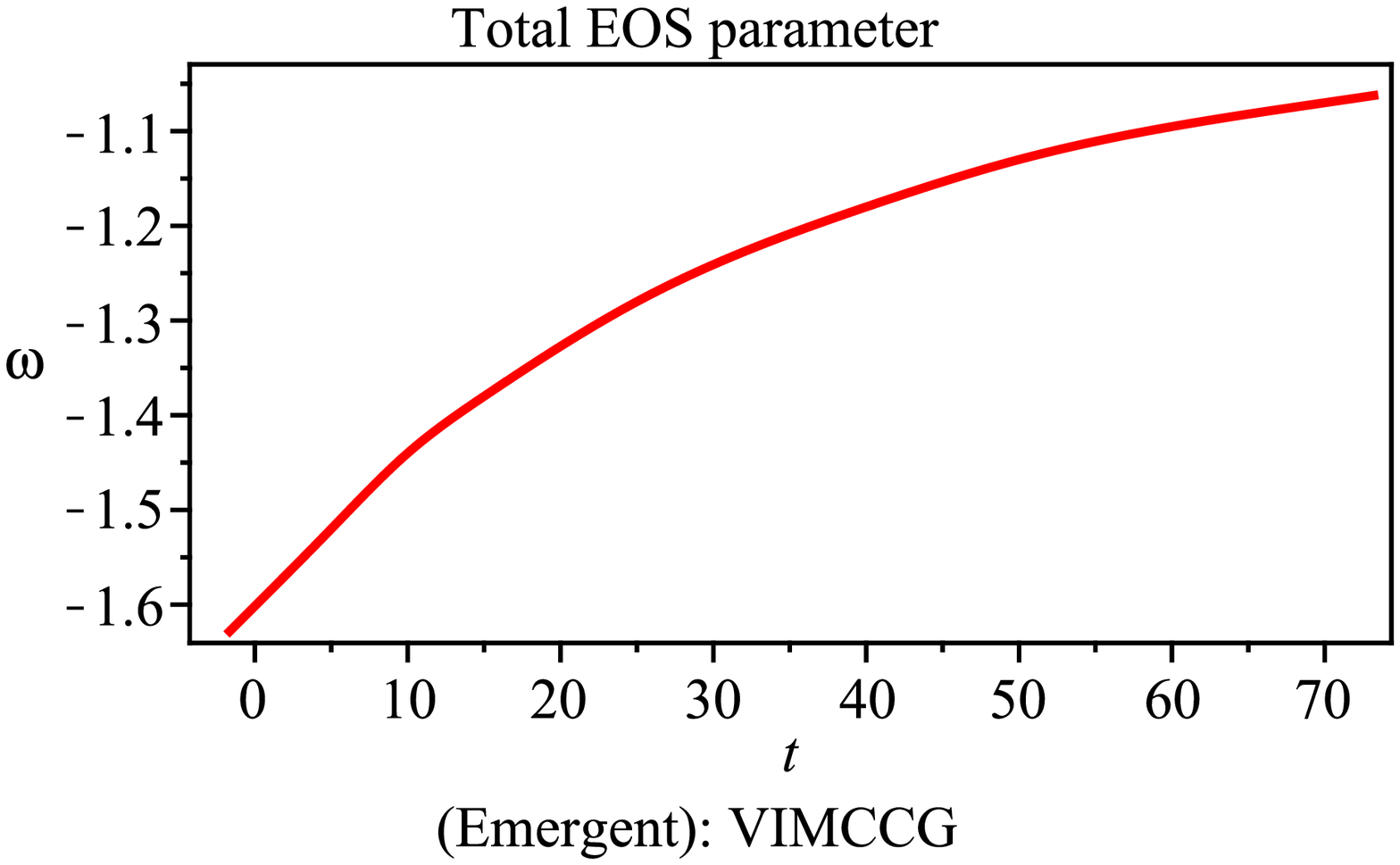}
\caption{Plot of $V$ and $\omega_{tot}$ in terms of time with
$m=1.1$, $A=0.1$, $b=1$, $K=0.03$, $\gamma=0.7$, $B=1$, $a_{0}=0.6$,
$\mu=0.3$, $\varsigma=1$, $\omega=-0.5$, $\alpha=0.5$ and $K=0.03$.}
\end{center}
\end{figure}

\begin{figure}[th]
\begin{center}
\includegraphics[scale=.4]{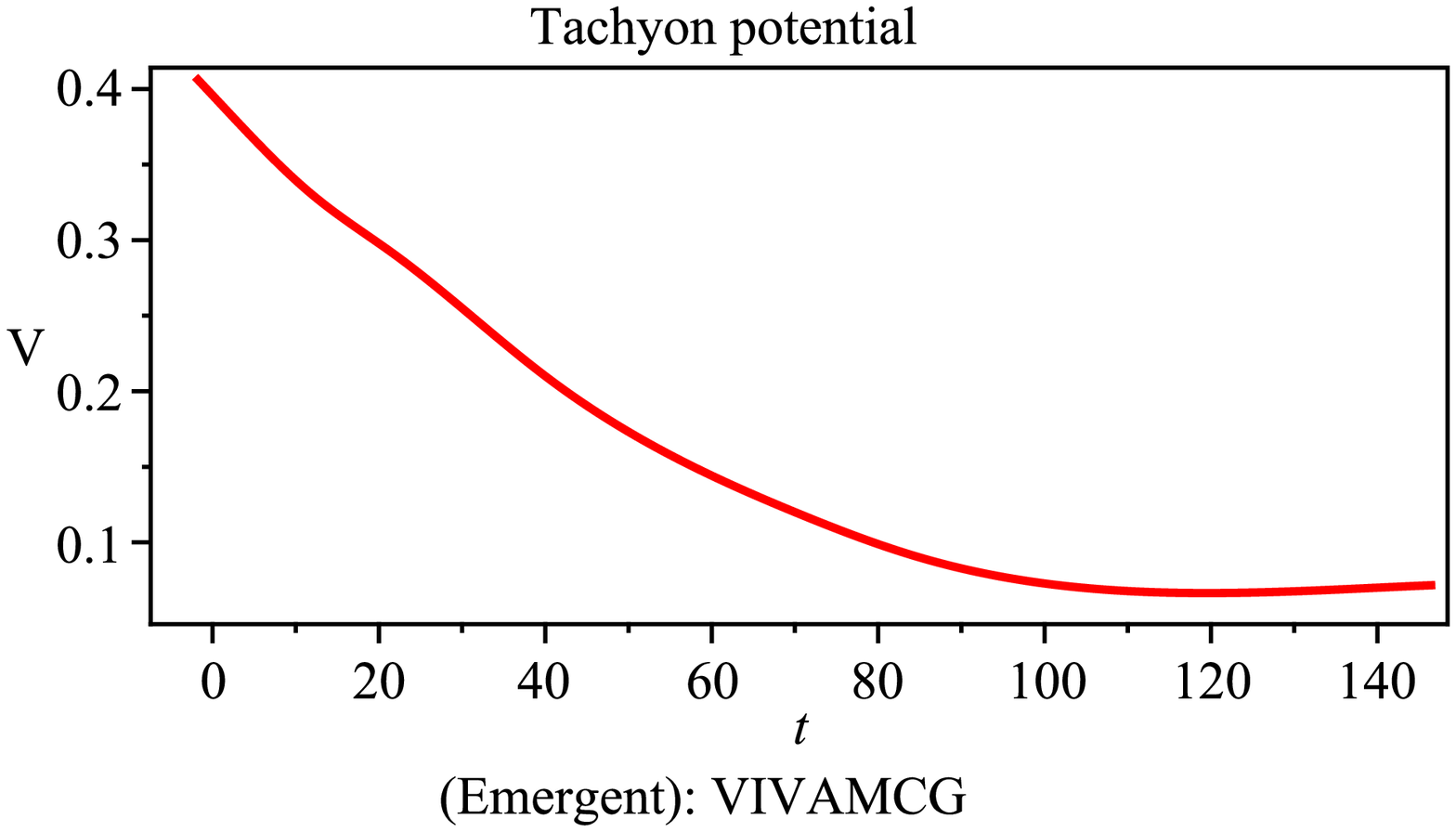}\includegraphics[scale=.4]{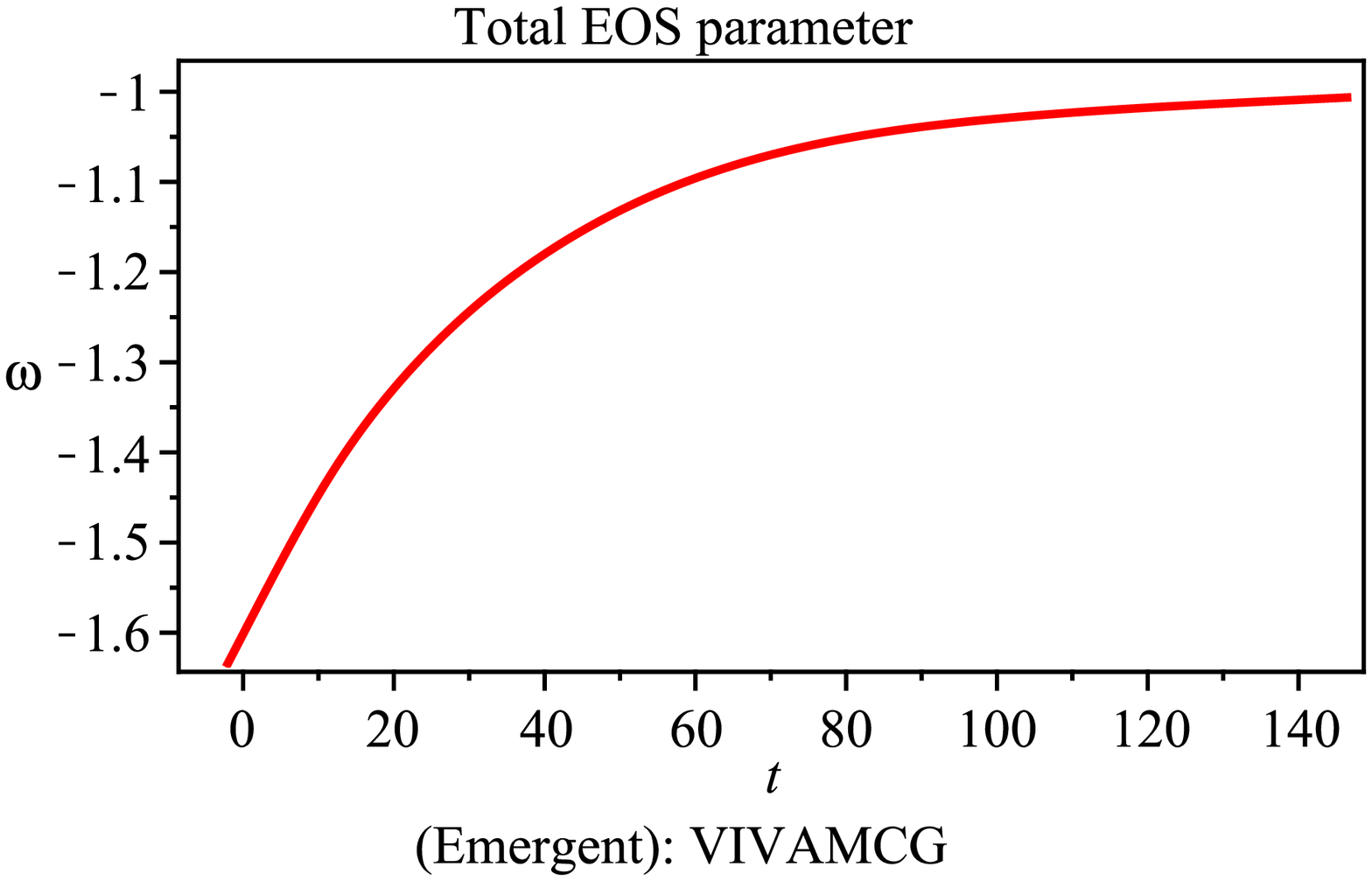}
\caption{Plot of $V$ and $\omega_{tot}$ in terms of time with
$m=1.1$, $b=0.1$, $K=0.03$, $\gamma=0.7$, $B=1$, $a_{0}=0.6$,
$i=-0.6$, $j=-0.08$, $\mu=0.3$, $\varsigma=1$, $\alpha=0.5$ and
$K=0.03$.}
\end{center}
\end{figure}

\begin{figure}[th]
\begin{center}
\includegraphics[scale=.4]{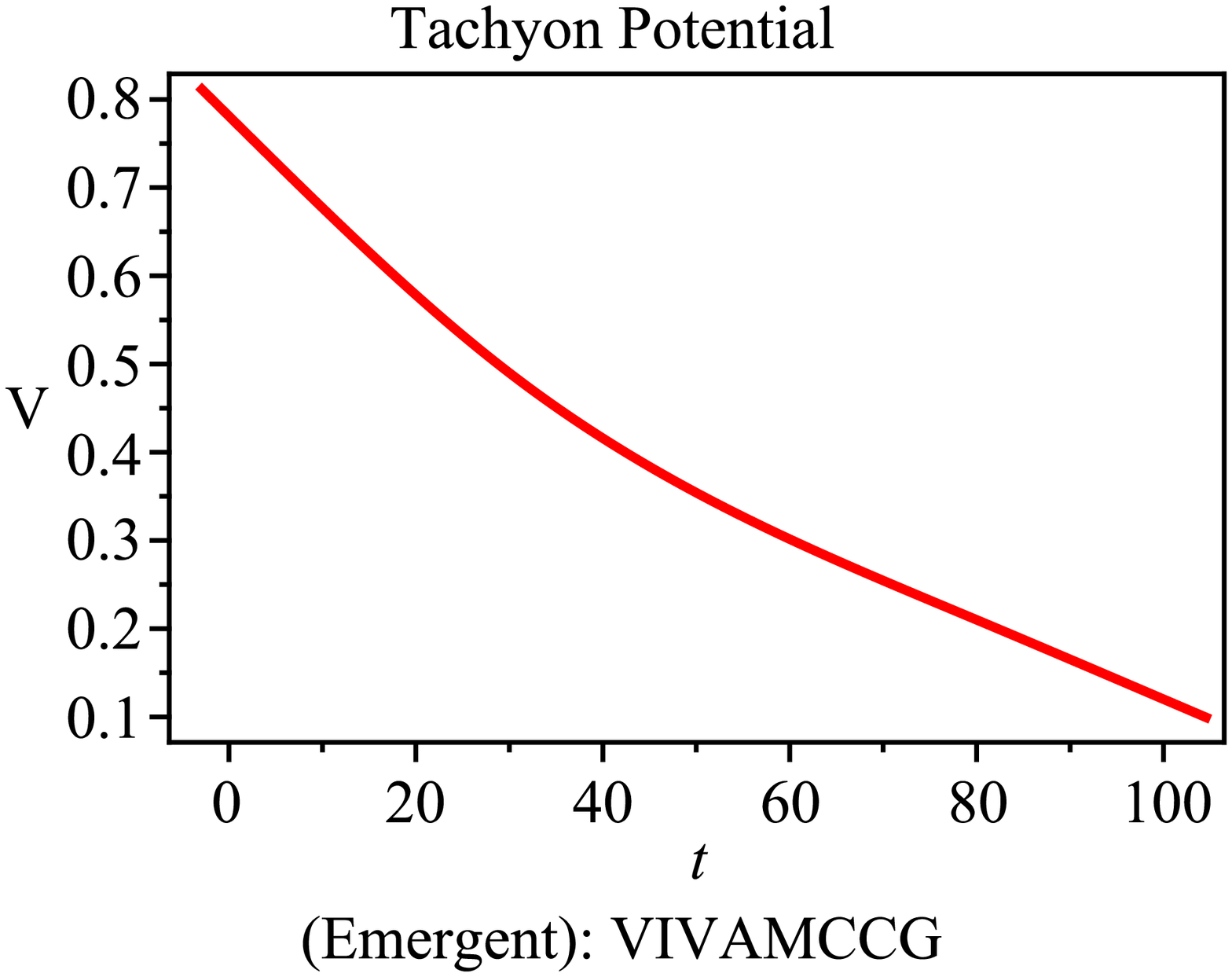}\includegraphics[scale=.4]{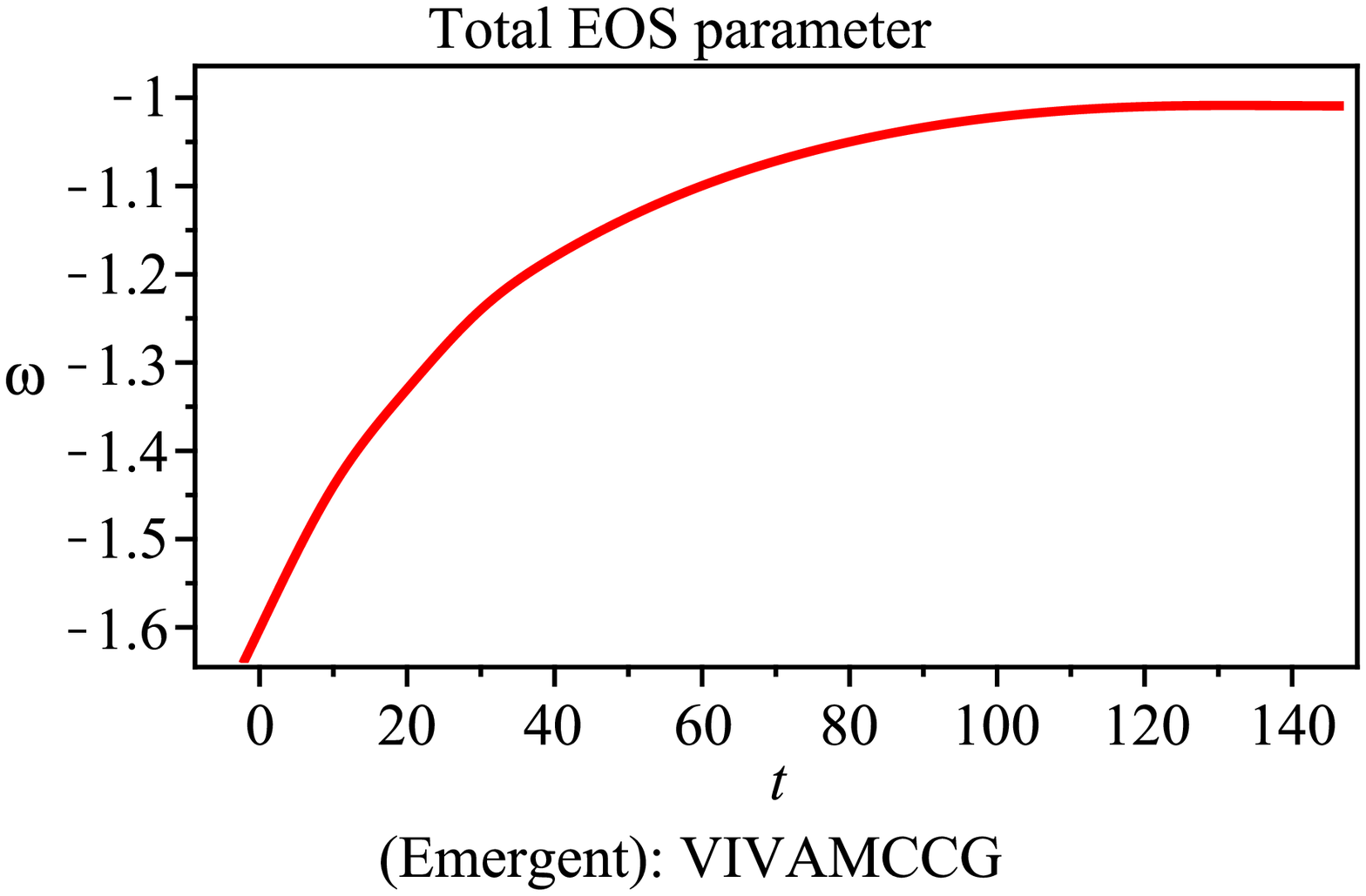}
\caption{Plot of $V$ and $\omega_{tot}$ in terms of time with
$m=1.1$, $b=1$, $K=0.03$, $\gamma=0.7$, $B=1$, $a_{0}=0.6$,
$i=-0.6$, $j=-0.8$, $\mu=0.3$, $\varsigma=1$, $\omega=-0.5$
$\alpha=0.5$ and $K=0.03$.}
\end{center}
\end{figure}

\begin{figure}[th]
\begin{center}
\includegraphics[scale=.4]{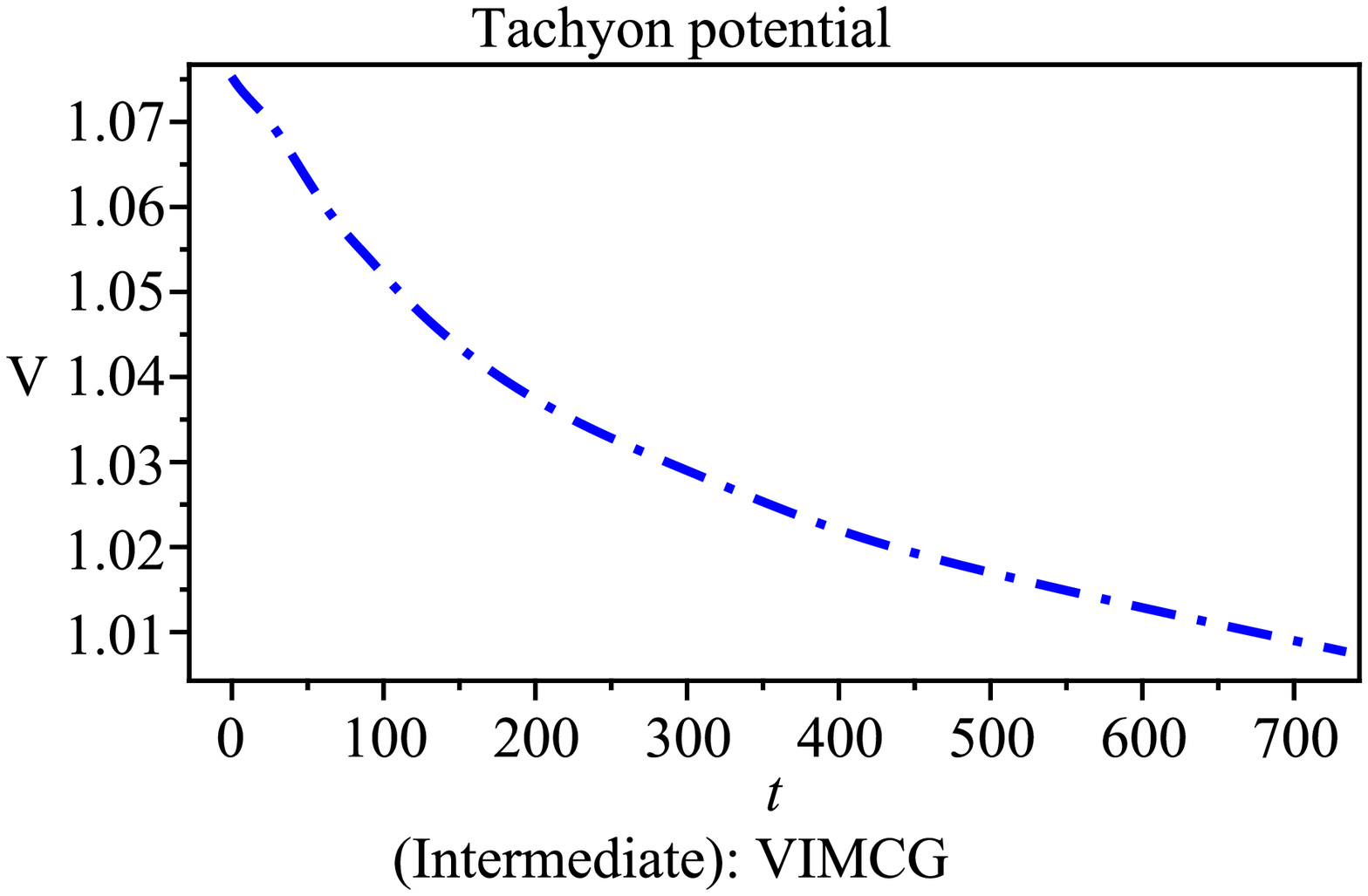}\includegraphics[scale=.4]{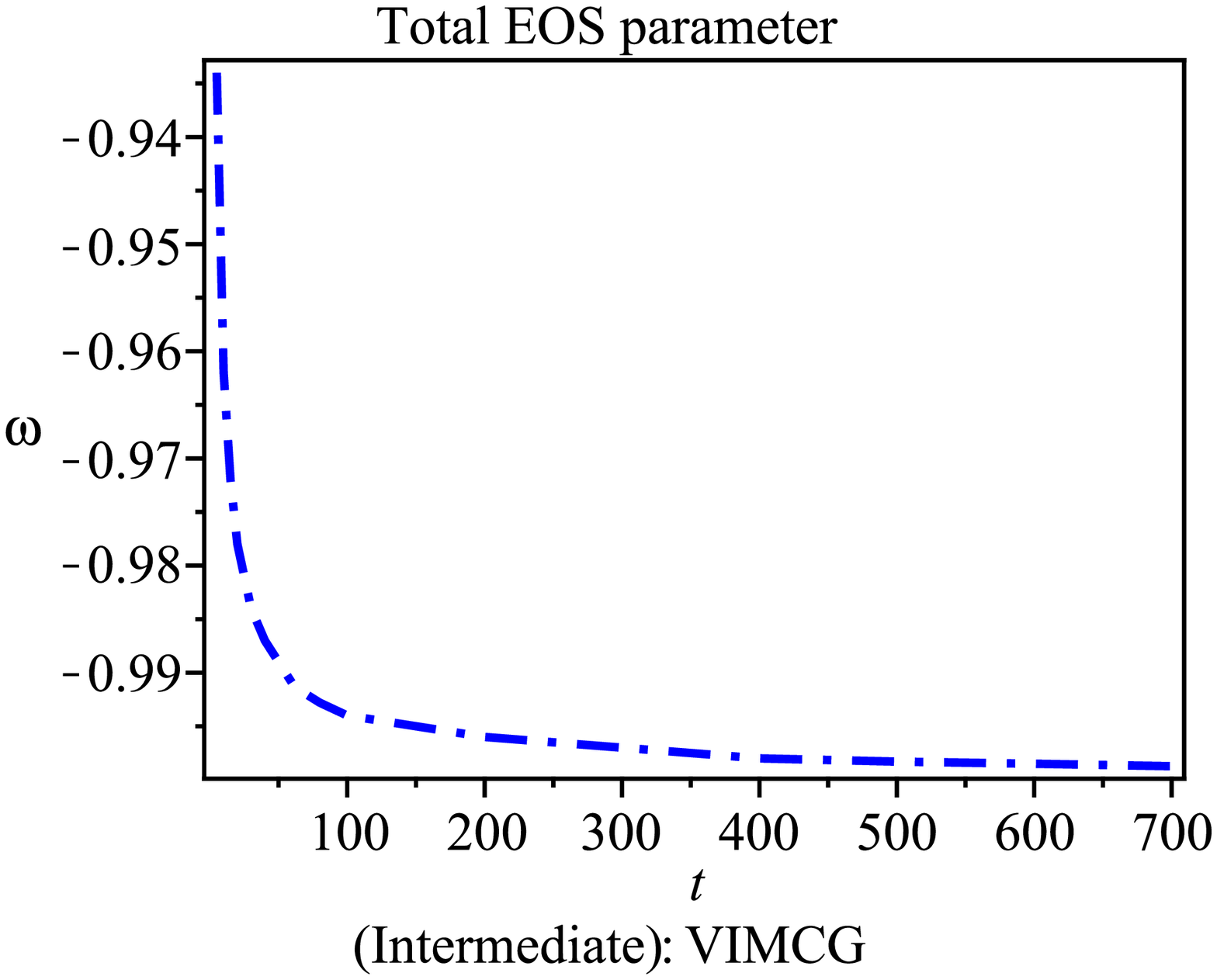}
\caption{Plot of $V$ and $\omega_{tot}$ in terms of time with
$\mu=0.3$, $\gamma=0.7$, $\lambda=0.7$, $\beta=0.8$, $\varsigma=1$,
$A=1$, $b=0.1$, and $\alpha=0.5$.}
\end{center}
\end{figure}

\begin{figure}[th]
\begin{center}
\includegraphics[scale=.4]{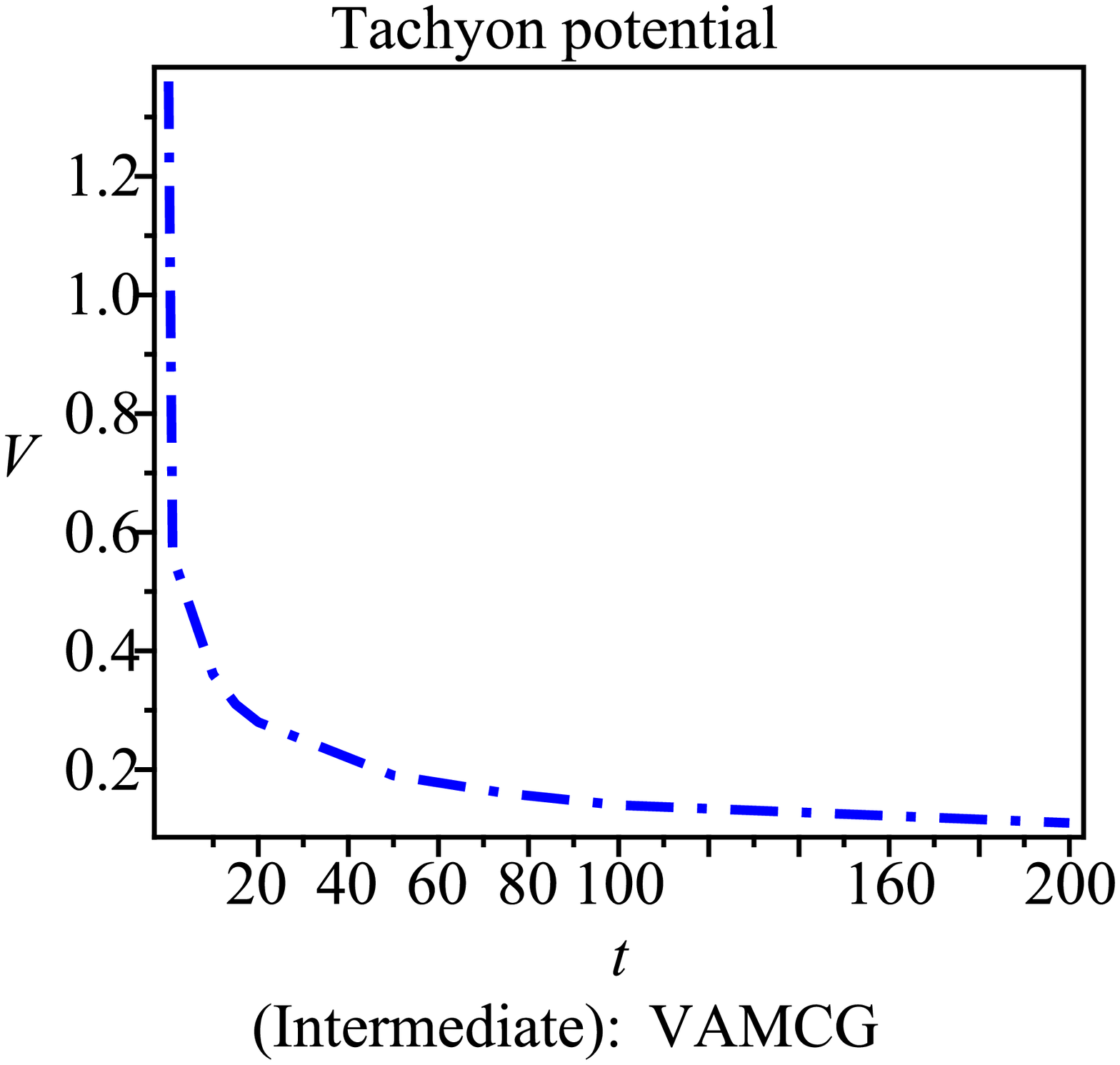}\includegraphics[scale=.4]{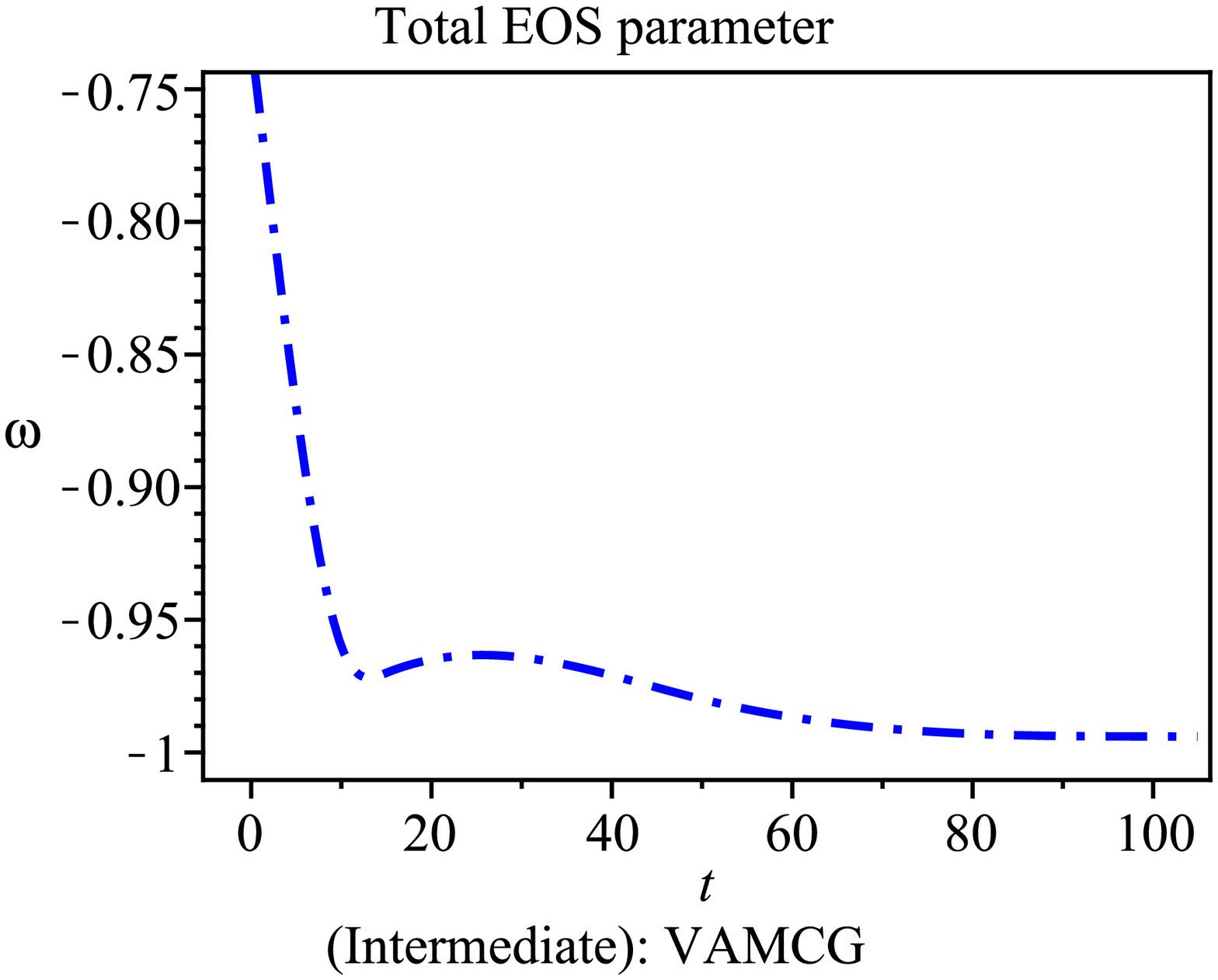}
\caption{Plot of $V$ and $\omega_{tot}$ in terms of time with
$\mu=0.3$, $\gamma=0.7$, $\lambda=0.7$, $\beta=0.8$, $b=0.1$,
$i=-0.6$, $j=-0.03$ and $\alpha=0.5$.}
\end{center}
\end{figure}

\begin{figure}[th]
\begin{center}
\includegraphics[scale=.4]{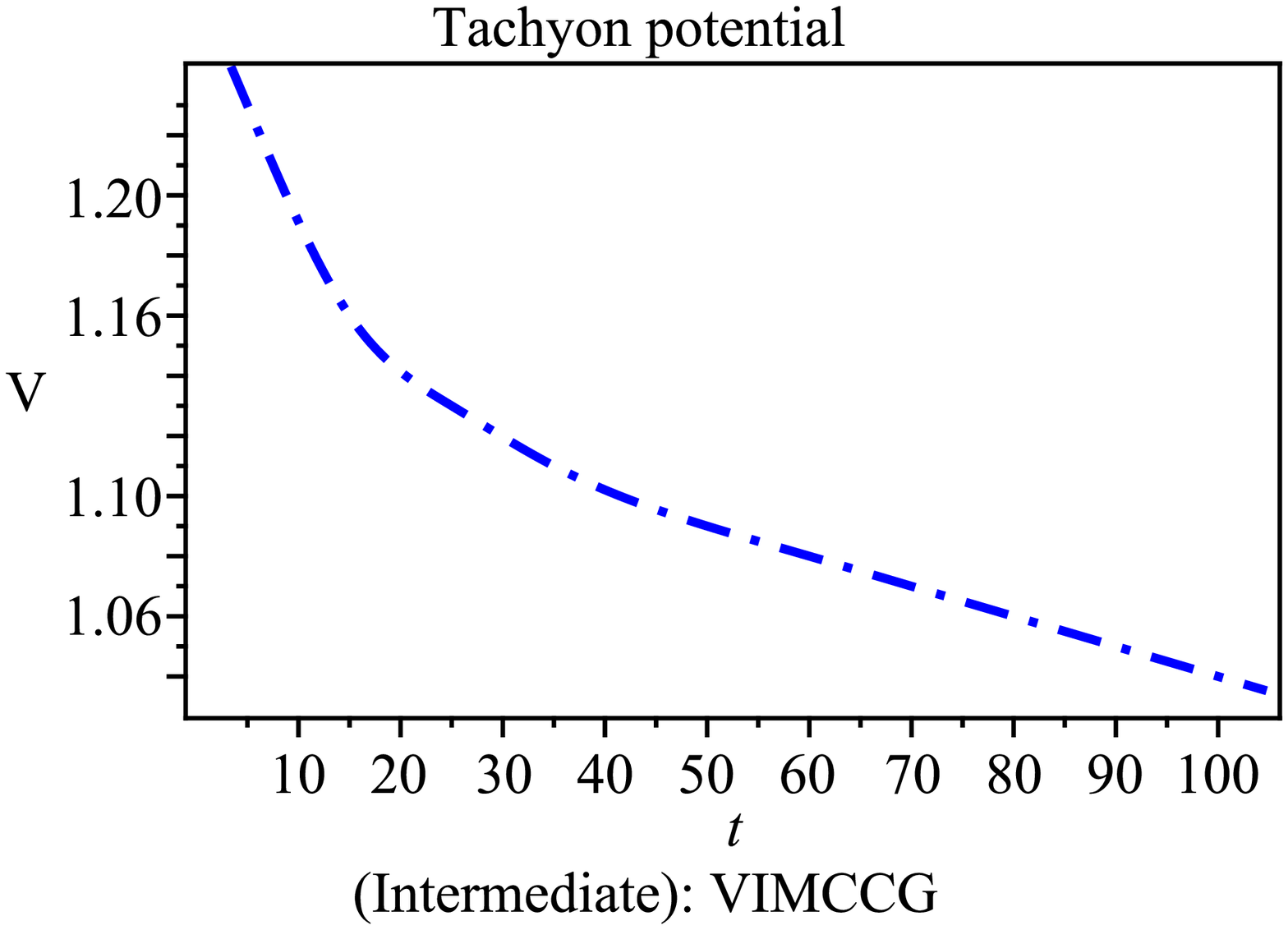}\includegraphics[scale=.4]{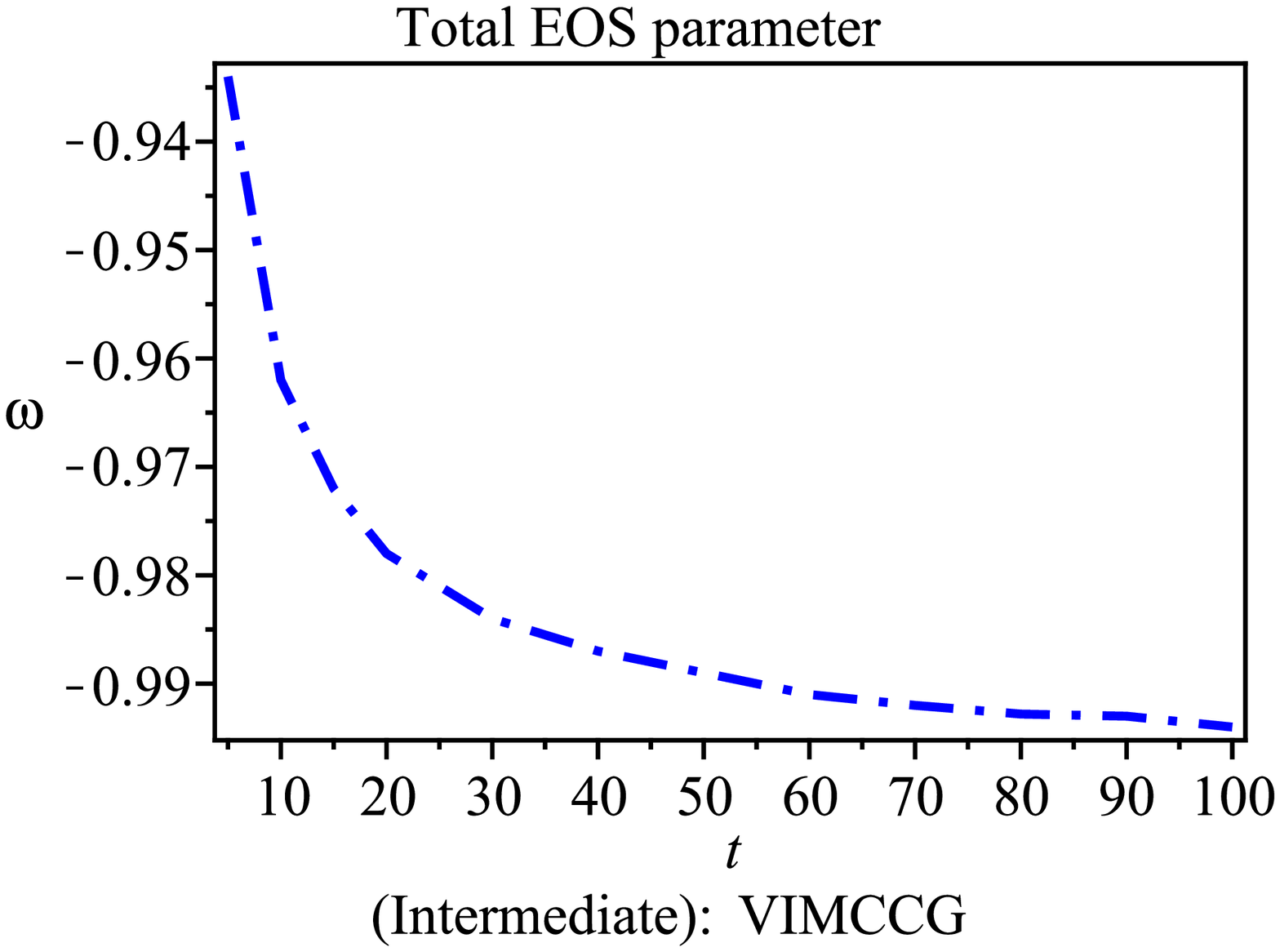}
\caption{Plot of $V$ and $\omega_{tot}$ in terms of time with
$\mu=0.3$, $\gamma=0.7$, $\lambda=0.7$, $\beta=0.8$, $\varsigma=1$,
$A=0.4$, $b=0.1$, $\omega=-0.5$ and $\alpha=0.5$}
\end{center}
\end{figure}

\begin{figure}[th]
\begin{center}
\includegraphics[scale=.4]{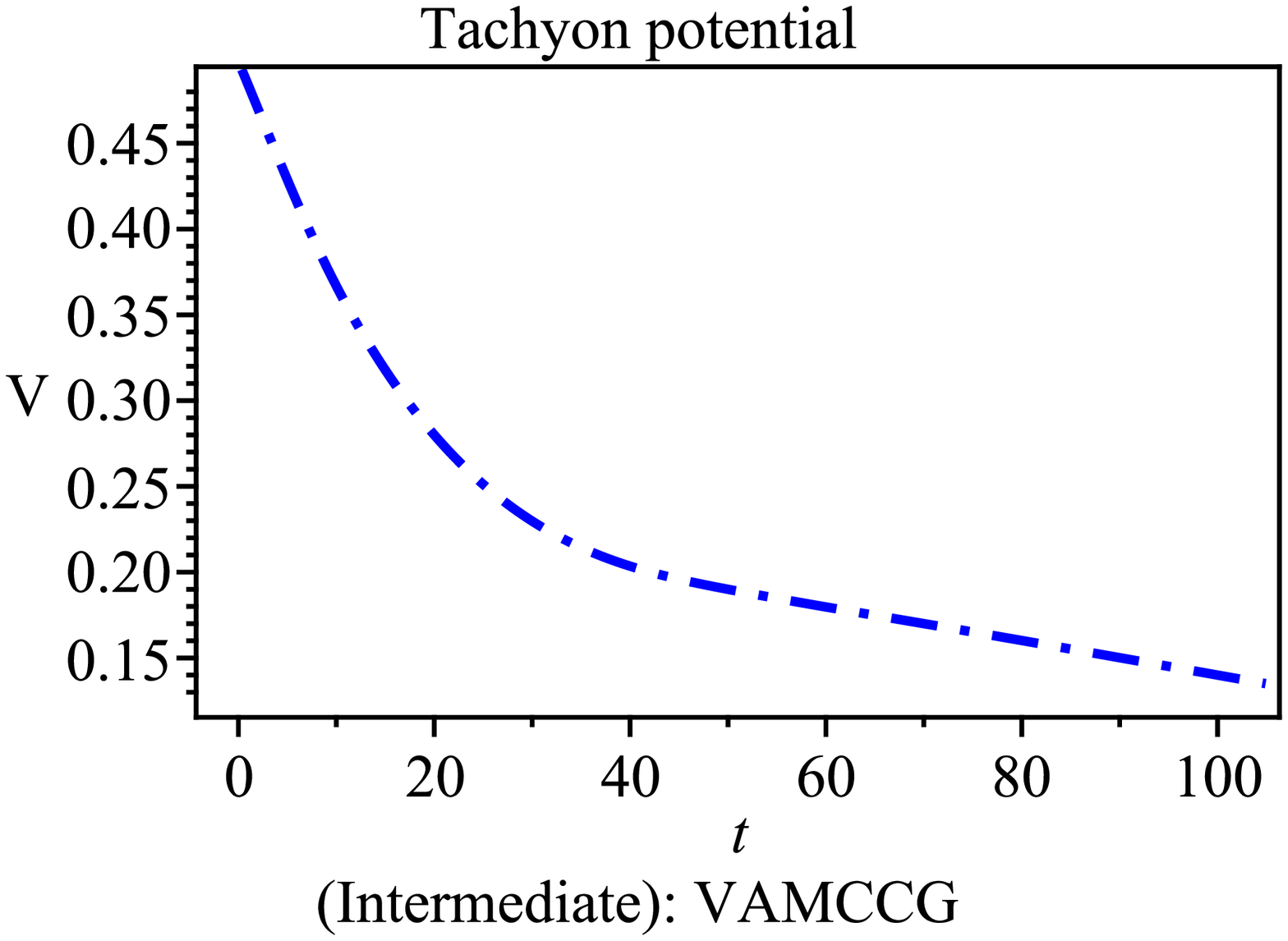}\includegraphics[scale=.4]{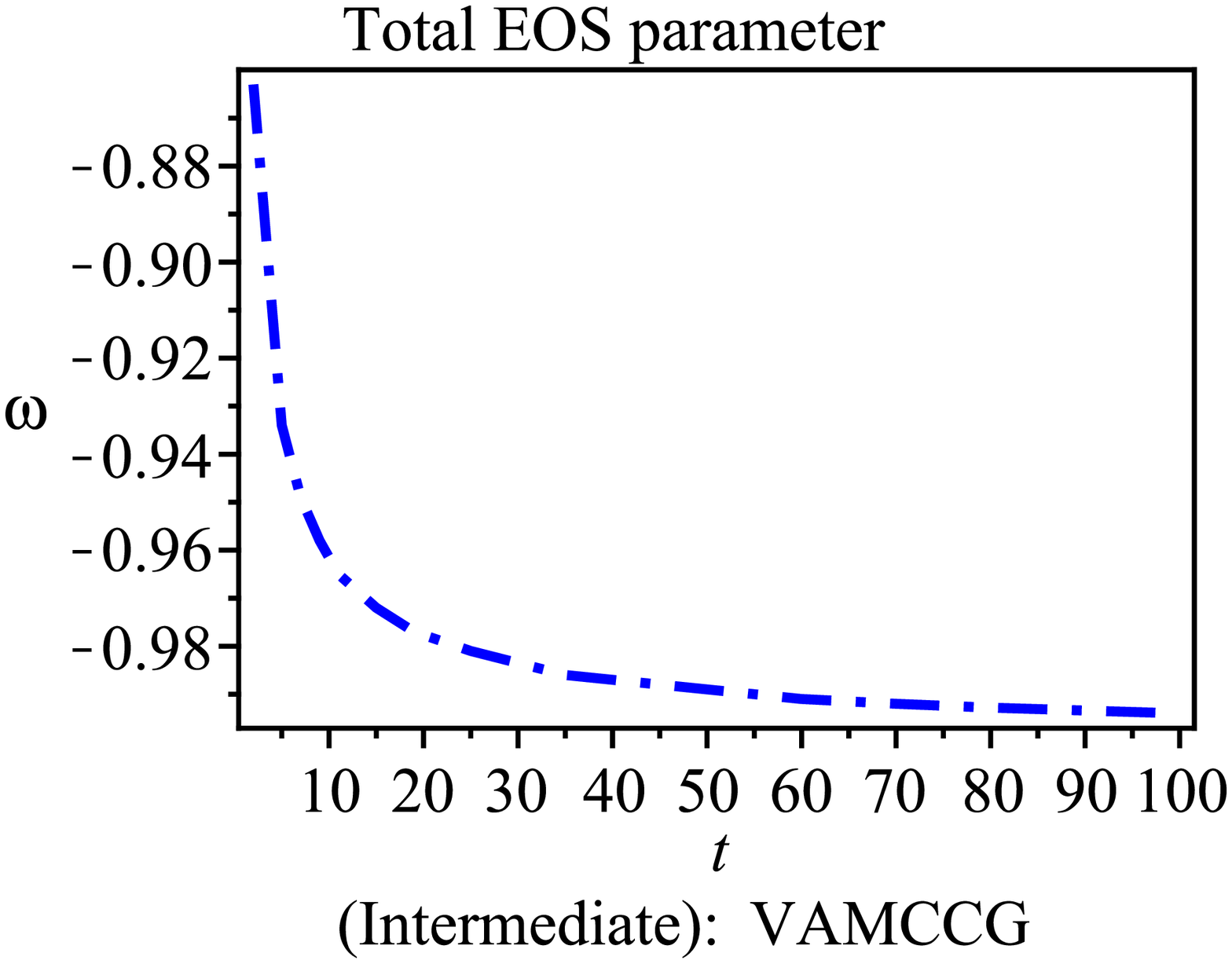}
\caption{Plot of $V$ and $\omega_{tot}$ in terms of time with
$\mu=0.3$, $\gamma=0.7$, $\lambda=0.7$, $\beta=0.8$, $\varsigma=1$,
$i=-0.6$, $j=-0.03$, $b=0.1$, $\omega=-0.5$, and $\alpha=0.5$.}
\end{center}
\end{figure}

\begin{figure}[th]
\begin{center}
\includegraphics[scale=.4]{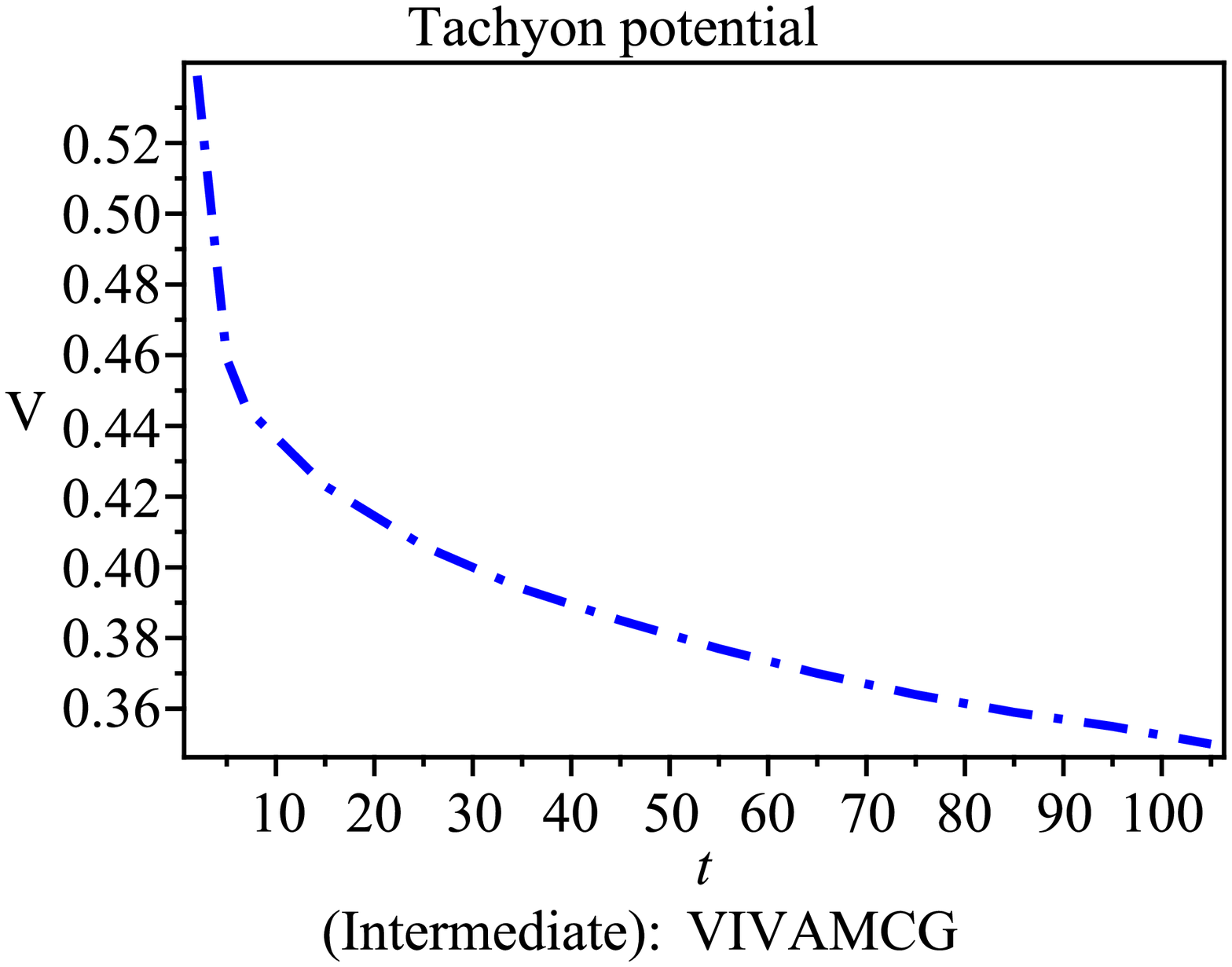}\includegraphics[scale=.4]{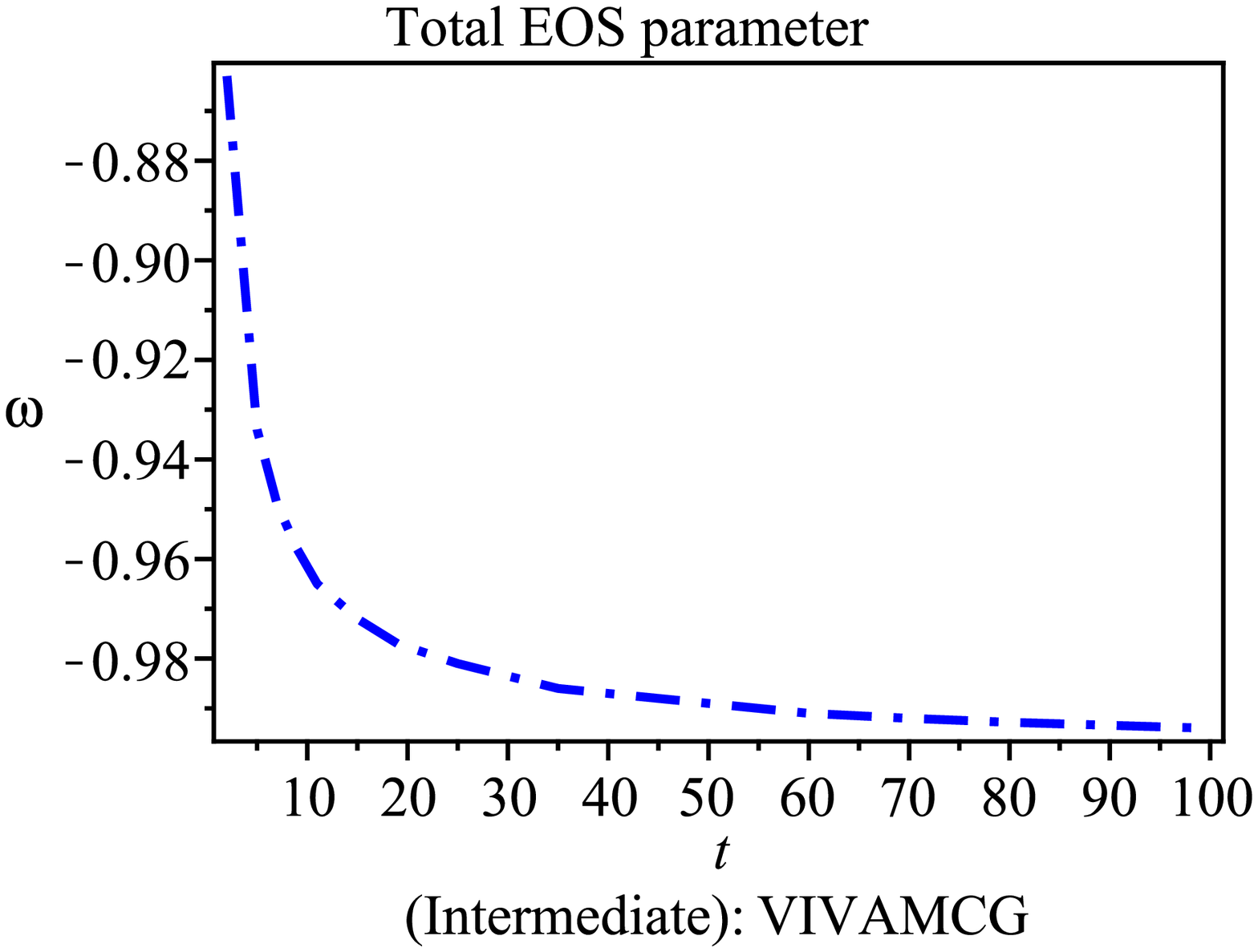}
\caption{Plot of $V$ and $\omega_{tot}$ in terms of time with
$\mu=0.3$, $\gamma=0.7$, $\lambda=0.7$, $\beta=0.8$, $\varsigma=1$,
$i=-0.6$, $j=-0.03$, $b=0.1$, and $\alpha=0.5$.}
\end{center}
\end{figure}

\begin{figure}[th]
\begin{center}
\includegraphics[scale=.4]{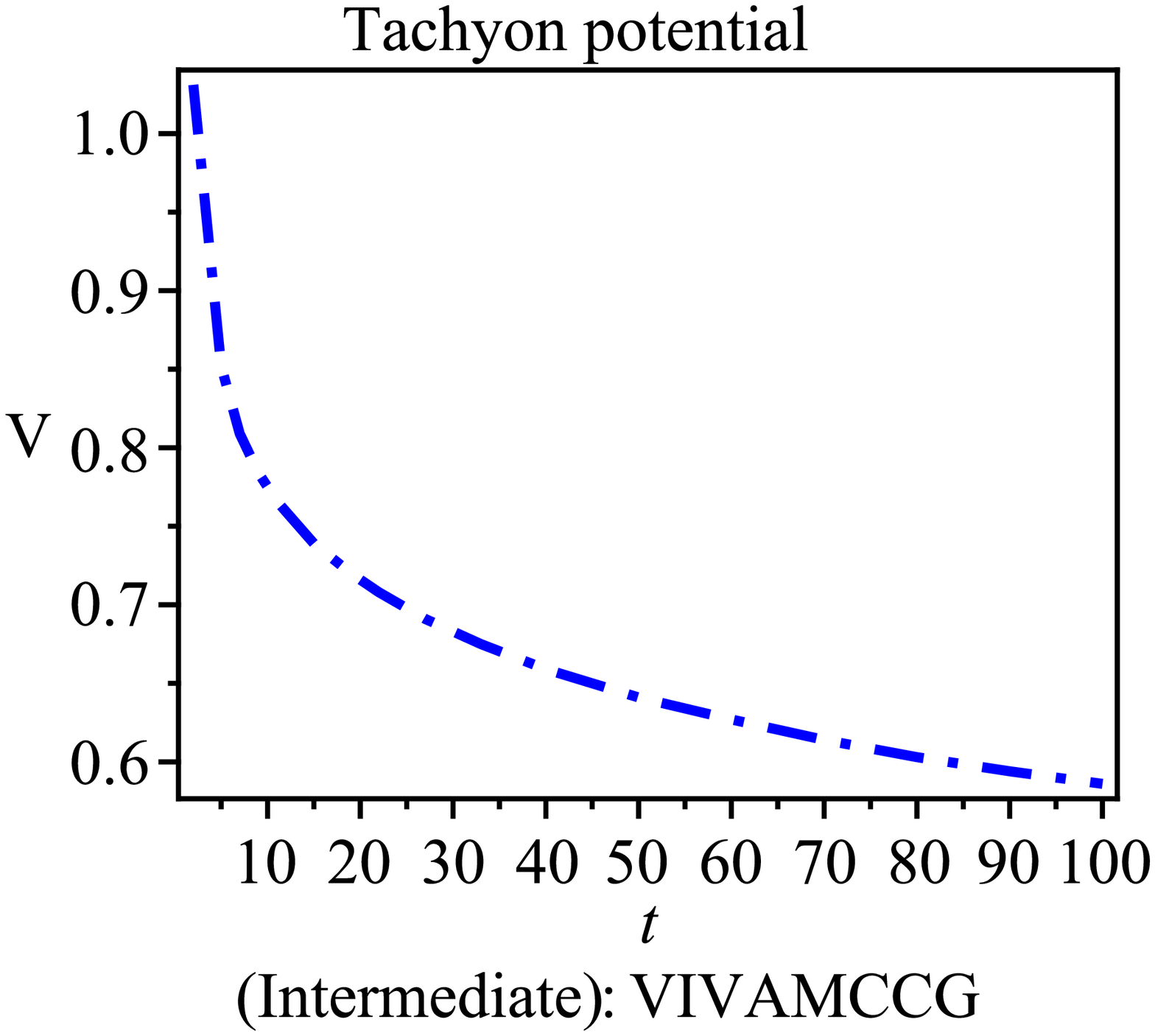}\includegraphics[scale=.4]{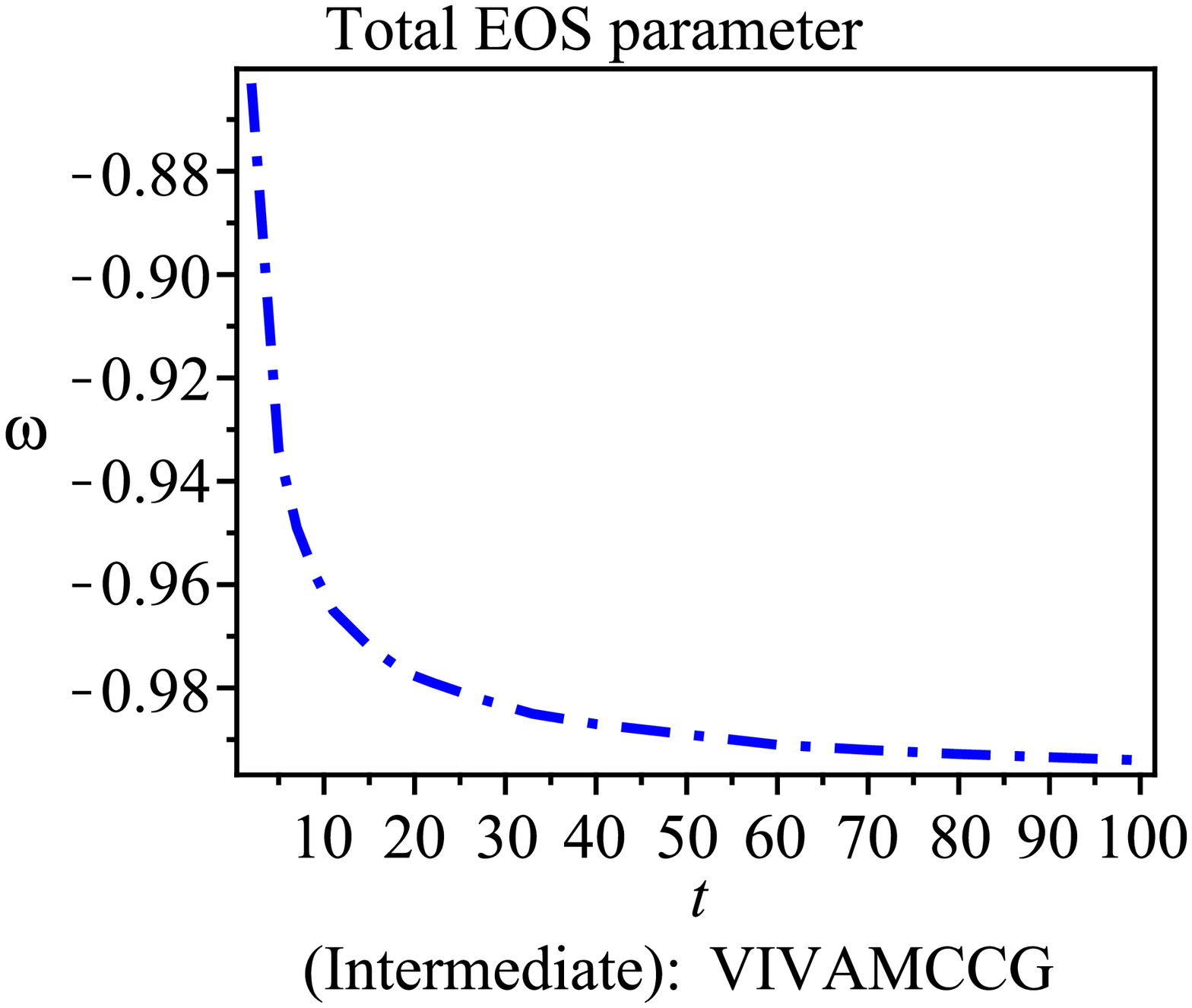}
\caption{Plot of $V$ and $\omega_{tot}$ in terms of time with
$\mu=0.3$, $\gamma=0.7$, $\lambda=0.7$, $\beta=0.8$, $\varsigma=1$,
$i=-0.6$, $j=-0.03$, $b=0.1$, $\omega=-0.5$, and $\alpha=0.5$.}
\end{center}
\end{figure}

\begin{figure}[th]
\begin{center}
\includegraphics[scale=.4]{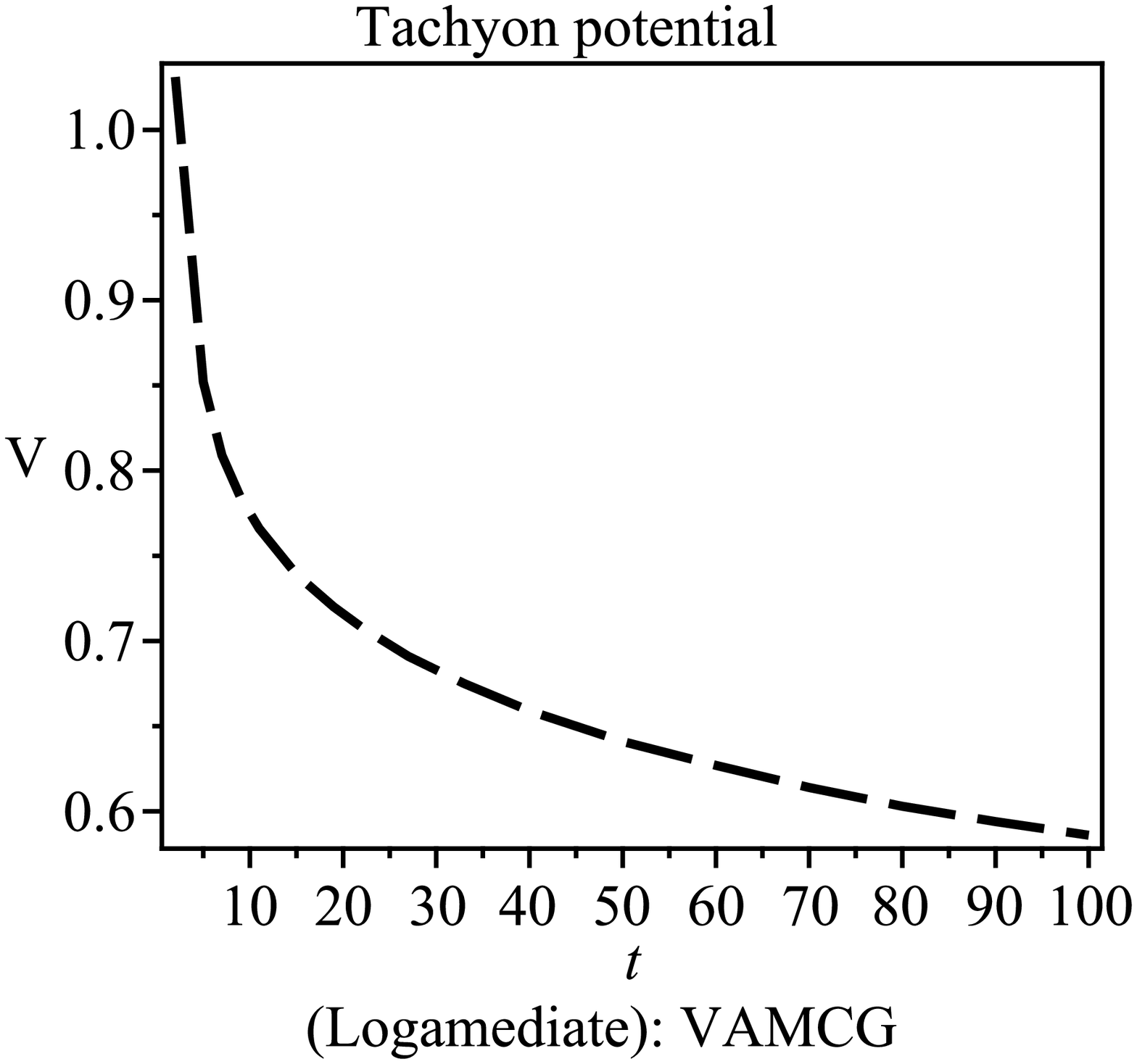}\includegraphics[scale=.4]{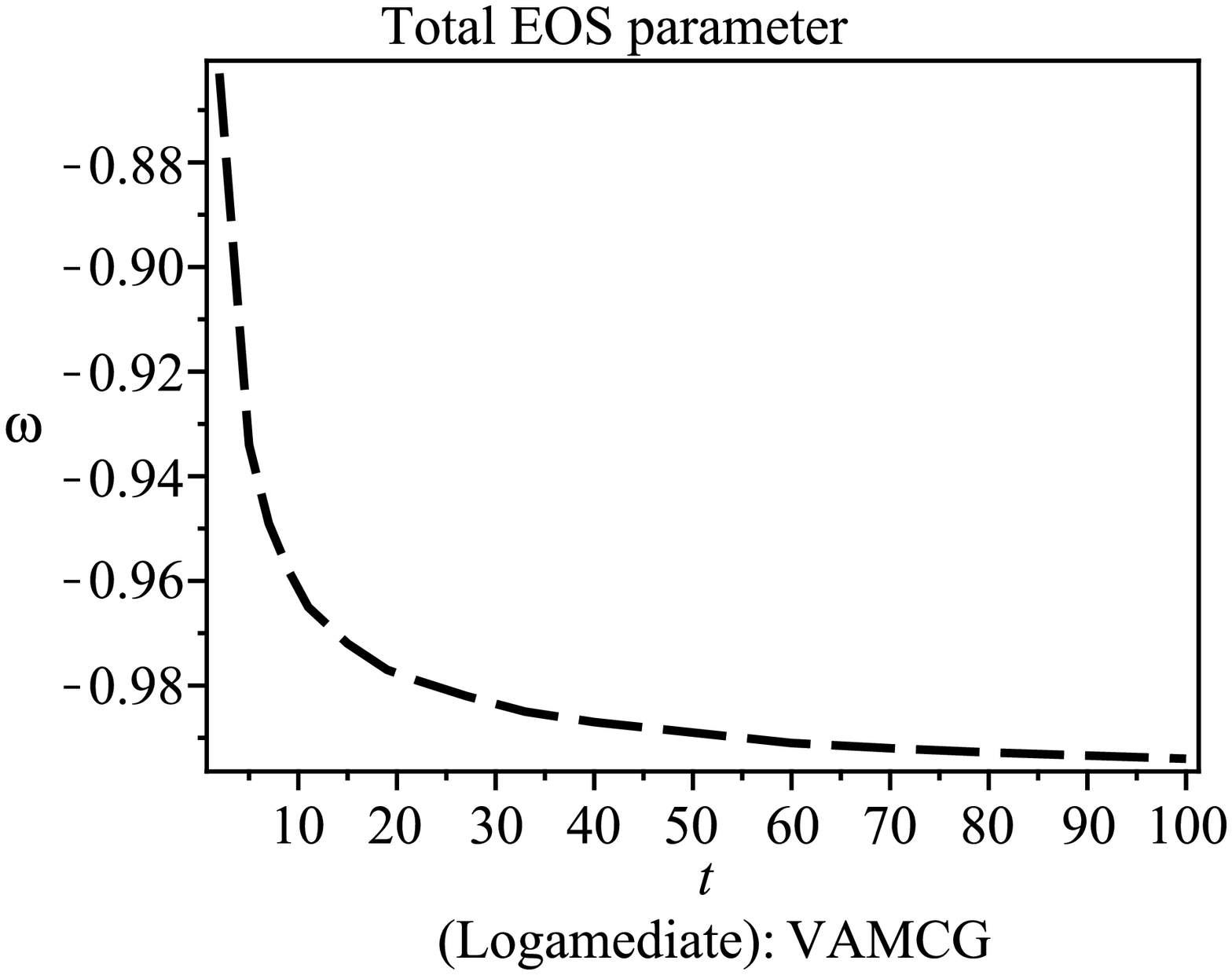}
\caption{Plot of $V$ and $\omega_{tot}$ in terms of time with
$x=0.05$, $b=0.4$, $\beta=2$, $\gamma=0.1$, $\alpha=0.5$, $\mu=0.3$,
$i=-0.8$ and $j=-0.4$.}
\end{center}
\end{figure}

\begin{figure}[th]
\begin{center}
\includegraphics[scale=.43]{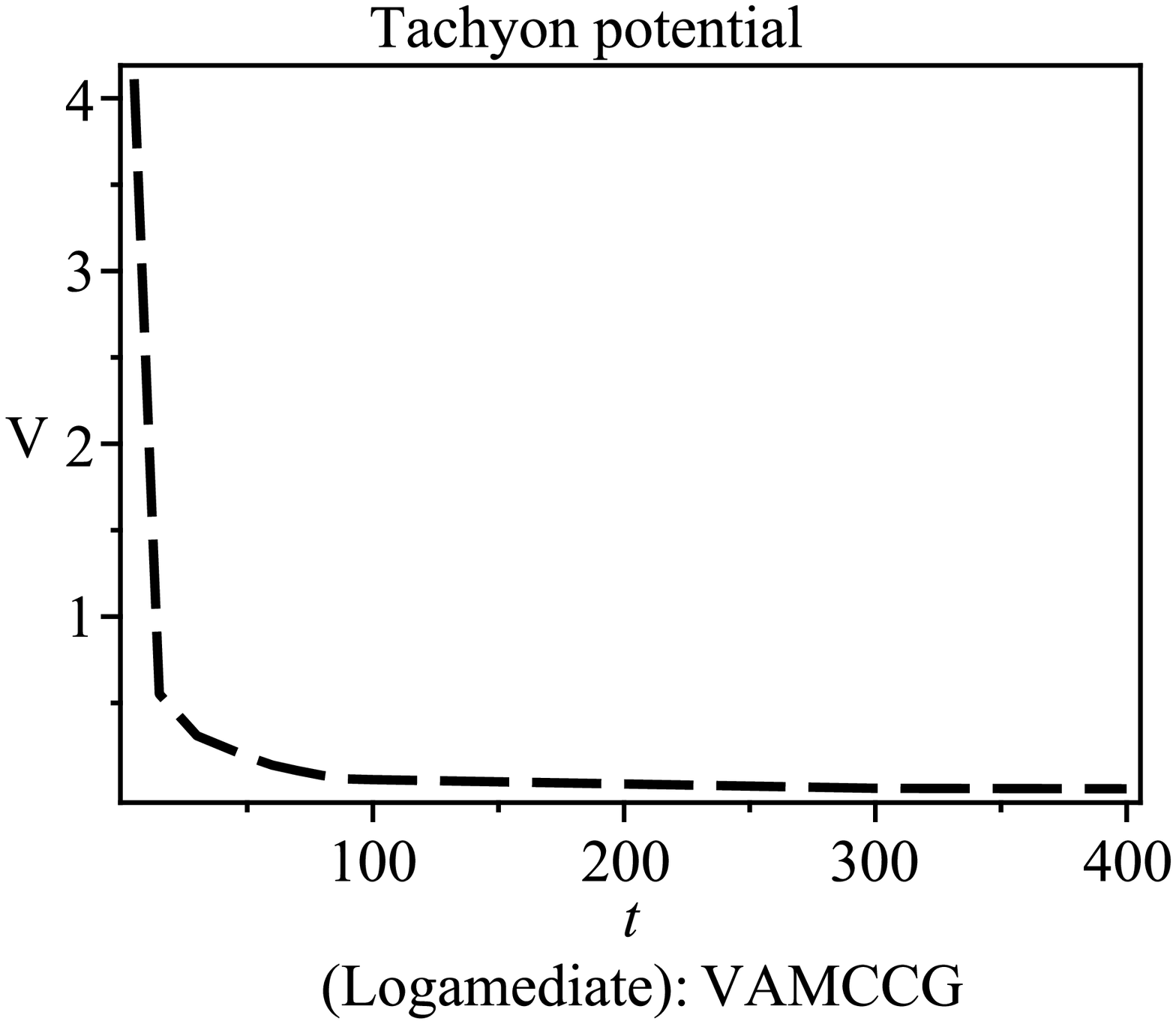}\includegraphics[scale=.4]{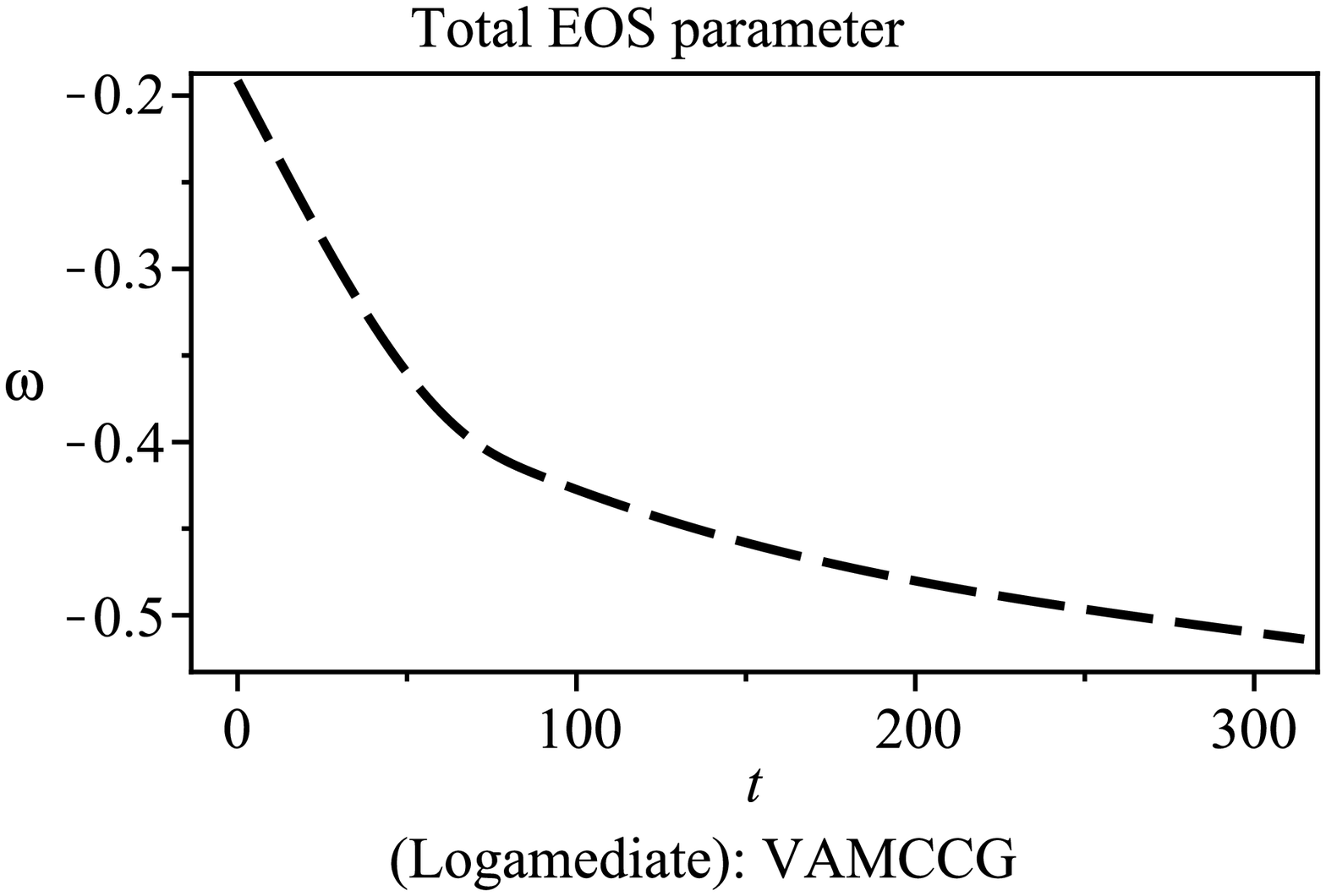}
\caption{Plot of $V$ and $\omega_{tot}$ in terms of time with
$x=0.1$, $b=0.4$, $\beta=2$, $\gamma=0.1$, $\alpha=0.5$, $\mu=0.3$,
$\omega=-0.5$, $i=-0.8$ and$j=-0.4$.}
\end{center}
\end{figure}

\begin{figure}[th]
\begin{center}
\includegraphics[scale=.4]{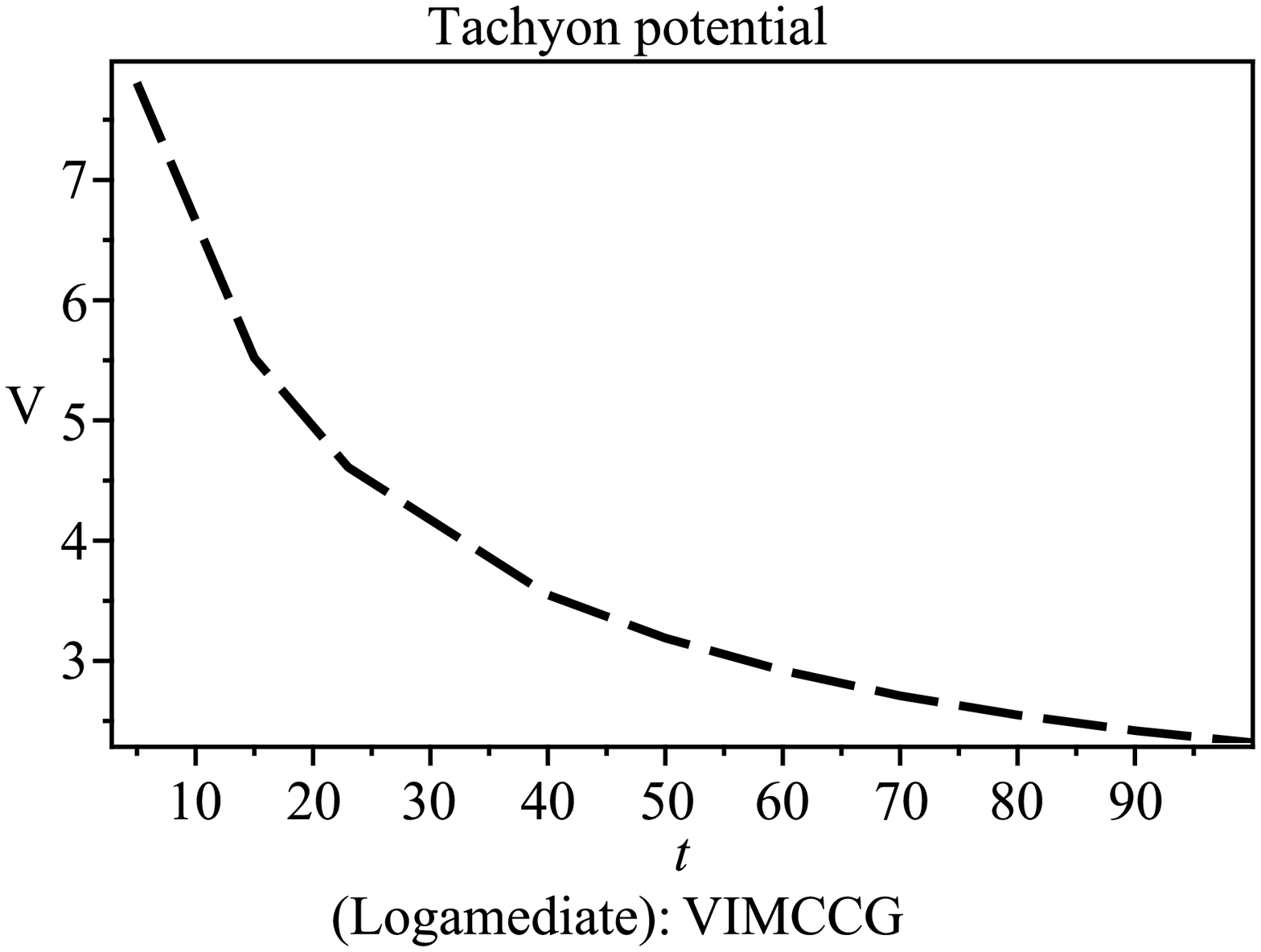}\includegraphics[scale=.4]{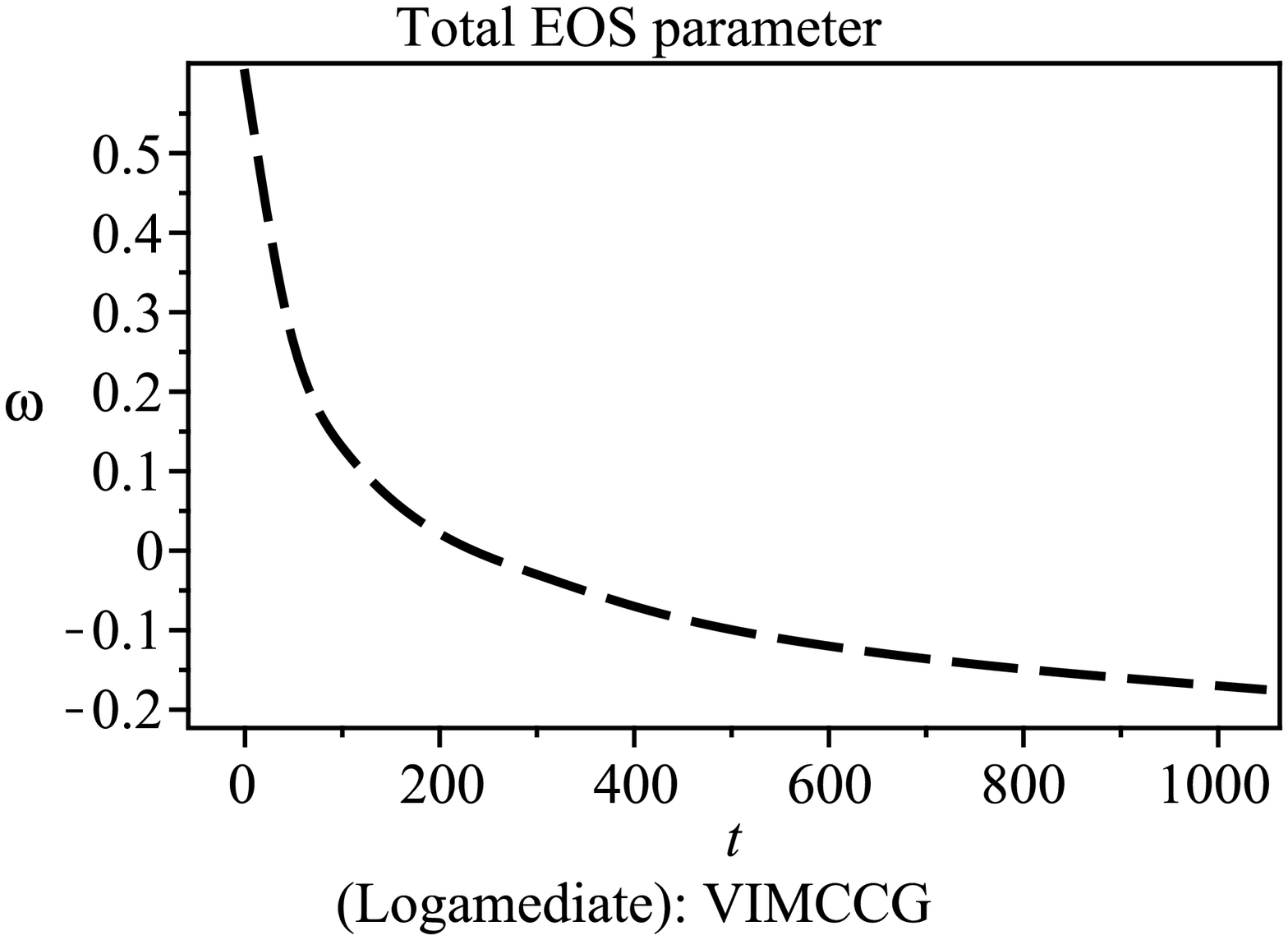}
\caption{Plot of $V$ and $\omega_{tot}$ in terms of time with
$x=0.05$, $b=0.4$, $\beta=2$, $\gamma=0.1$, $\alpha=0.5$, $\mu=0.3$,
$\varsigma=1$,$\omega=-0.5$ and $A=1$.}
\end{center}
\end{figure}

\begin{figure}[th]
\begin{center}
\includegraphics[scale=.4]{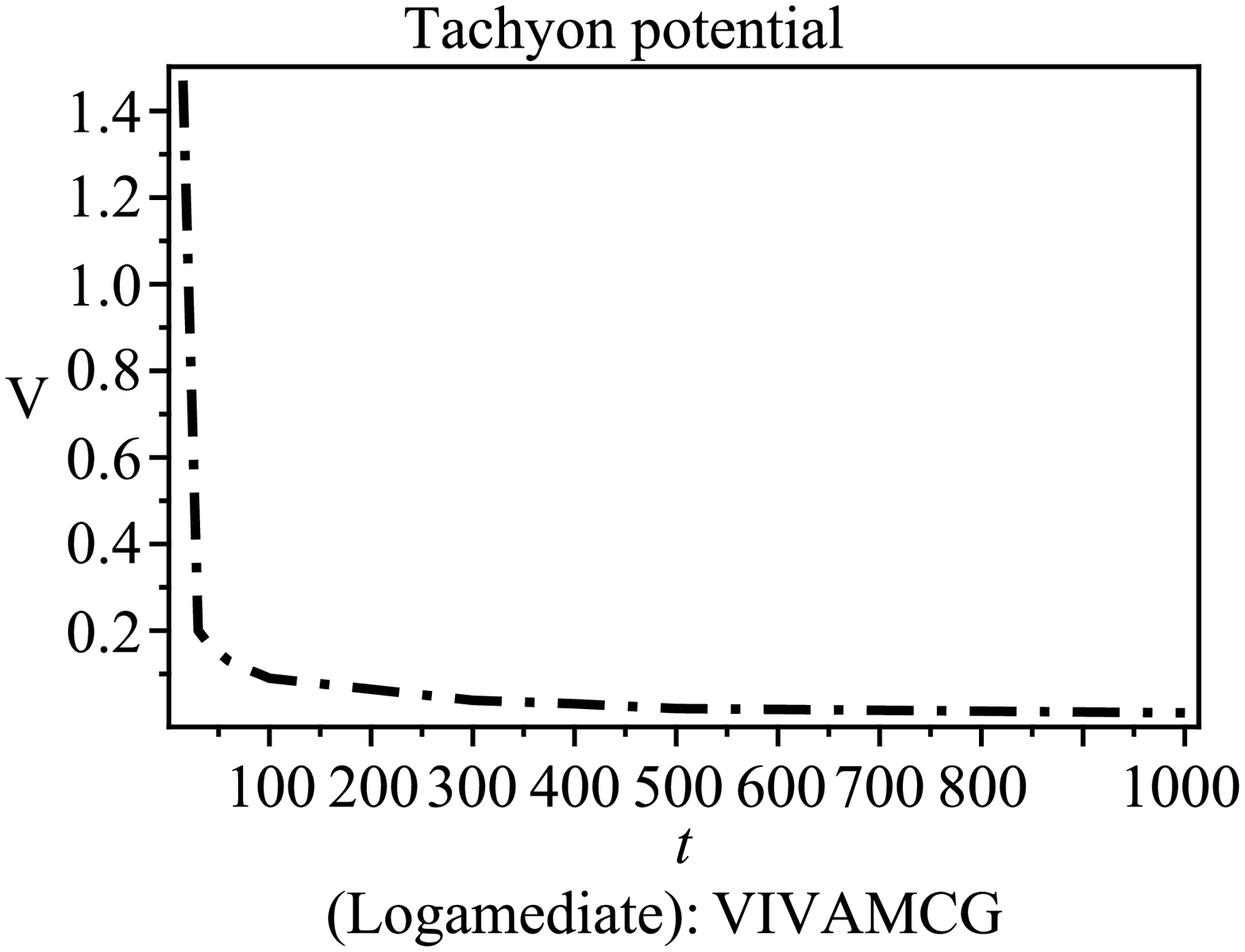}\includegraphics[scale=.4]{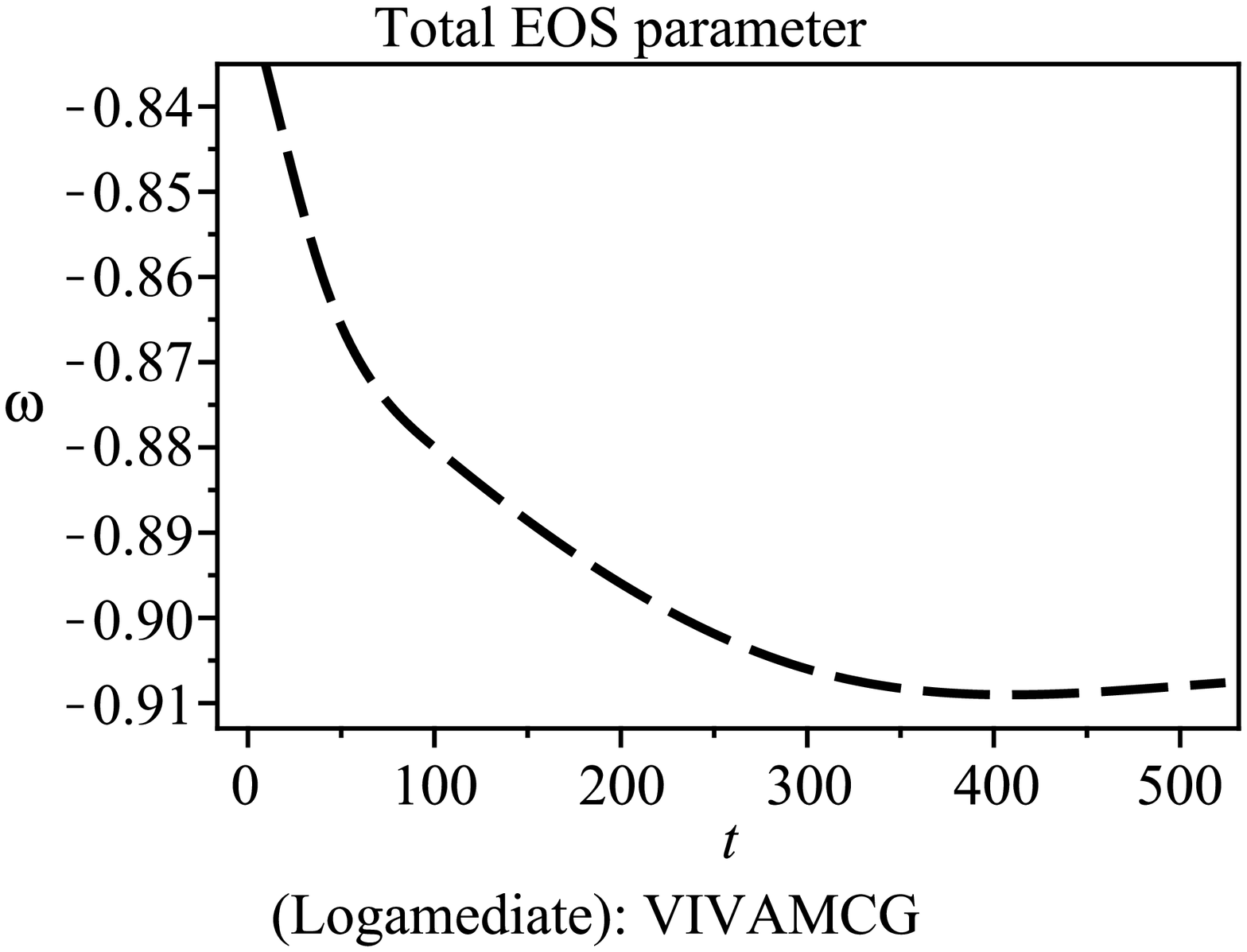}
\caption{Plot of $V$ and $\omega_{tot}$ in terms of time with
$x=0.5$, $b=0.4$, $\beta=2$, $\gamma=0.1$, $\alpha=0.5$, $\mu=0.3$,
$\varsigma=1$, $i=-0.8$ and $j=-0.7$.}
\end{center}
\end{figure}

\begin{figure}[th]
\begin{center}
\includegraphics[scale=.43]{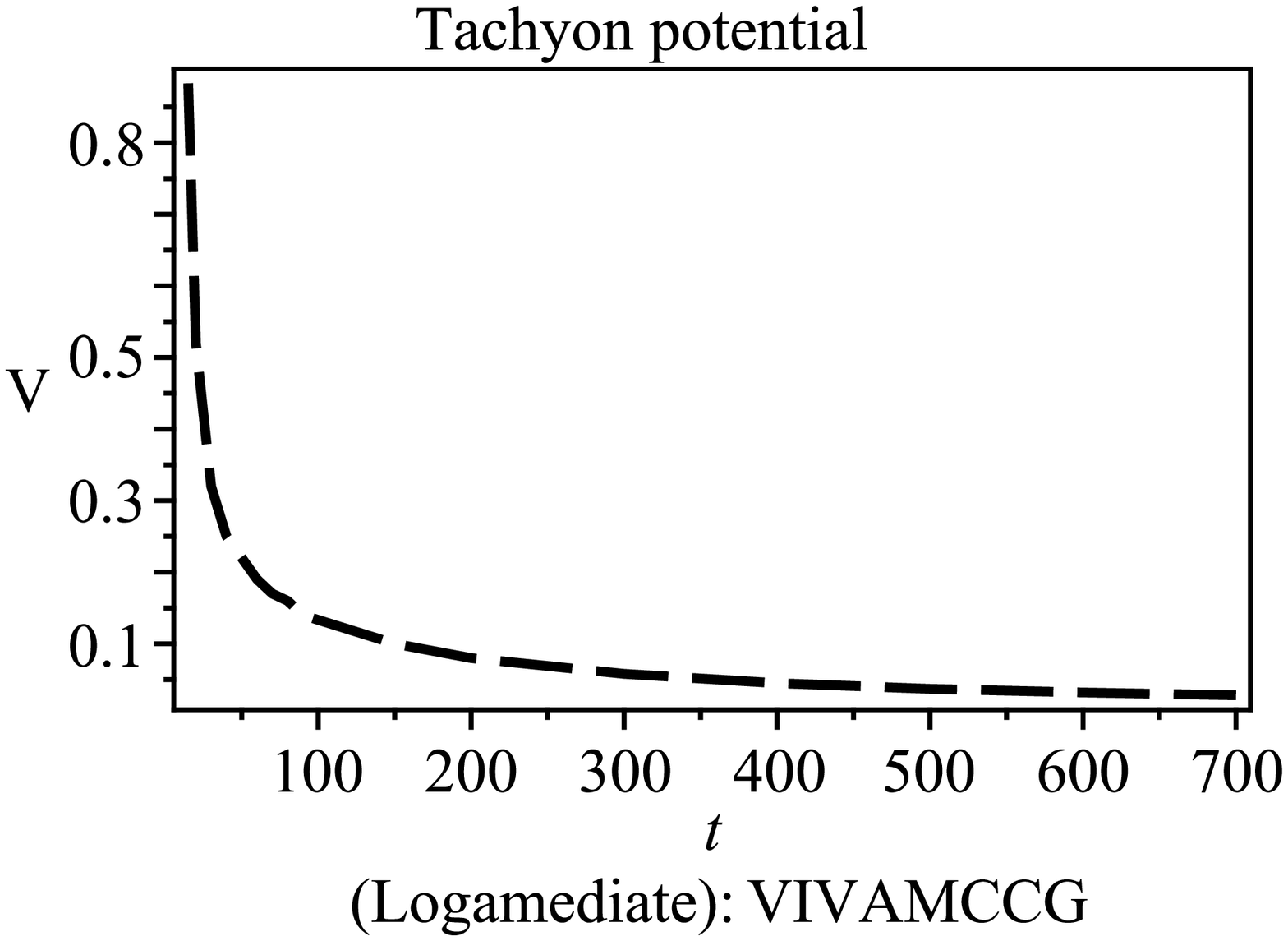}\includegraphics[scale=.4]{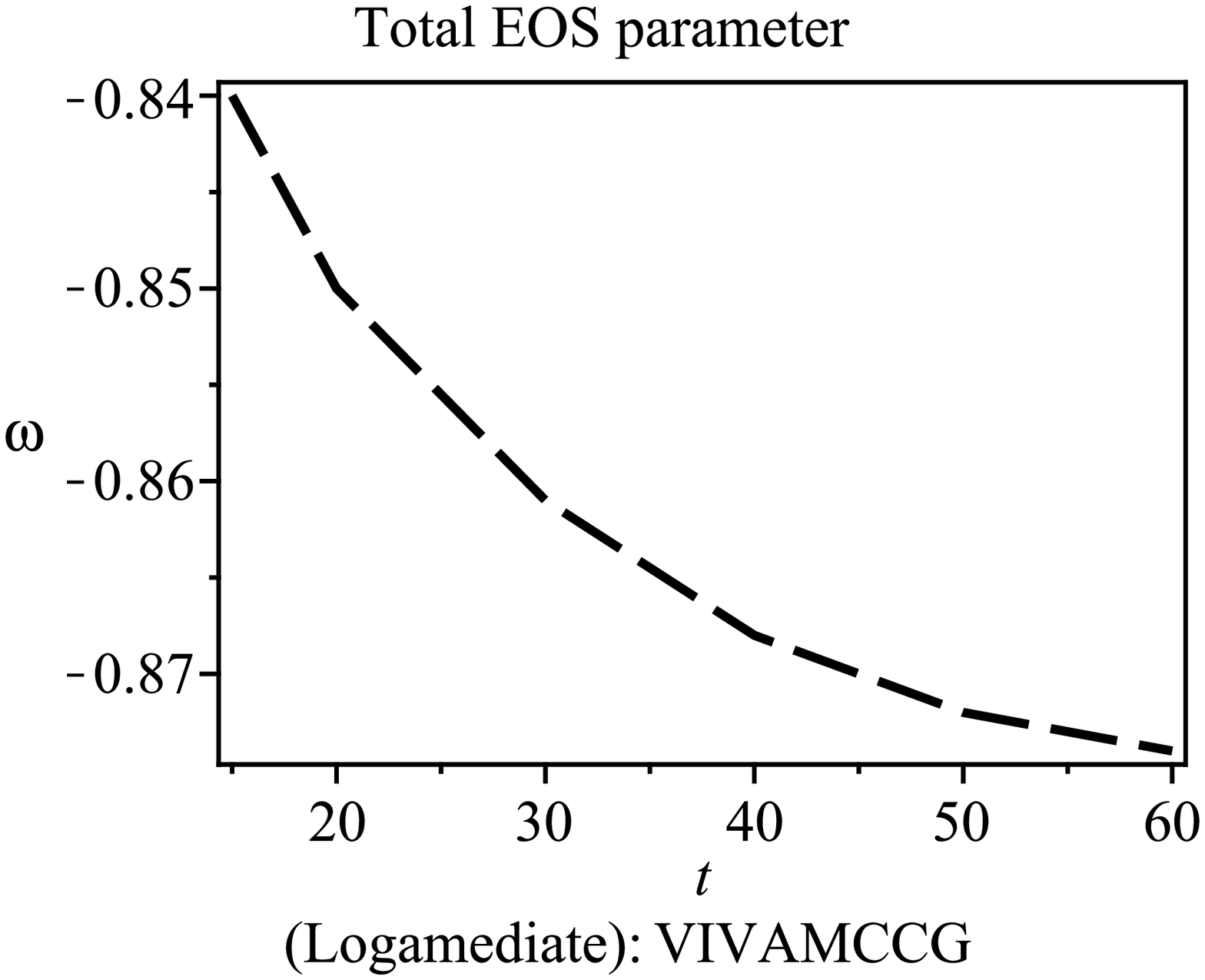}
\caption{Plot of $V$ and $\omega_{tot}$ in terms of time with
$x=0.5$, $b=0.4$, $\beta=2$, $\gamma=0.1$, $\alpha=0.5$, $\mu=0.3$,
$\varsigma=1$, $i=-0.8$, $j=-0.7$ and $\omega=-0.5$.}
\end{center}
\end{figure}

\end{document}